\documentclass{aa}

\usepackage{graphicx}
\usepackage{txfonts}
\usepackage{hyperref}
\usepackage{graphicx}   
\usepackage{amsmath}    
\usepackage{amssymb}    

\newcommand{\hi}{\textsc{H\,i}}

\newcommand{\kms}{\,$\rm km\,s^{-1}$}
\newcommand{\magasec}{\,mag\,arcsec$^{-2}$}
\newcommand{\msunpcsq}{\,$\rm M_{\odot}\,pc^{-2}$}

\newcommand{\logmstar}{$\log\,M_{\star}\,{\rm [M_{\odot}]}$}

\newcommand{\wifes}{\textsc{WiFeS}}

\newcommand{\hopcat}{\textsc{Hopcat}}
\newcommand{\hipass}{\textsc{Hipass}}

\newcommand{\highmass}{H\textsc{i}ghMass}
\newcommand{\bluedisk}{\textsc{Bluedisk}}
\newcommand{\hix}{\textsc{hix}}
\newcommand{\control}{\textsc{control}}

\newcommand{\pywifes}{\textsc{pywifes}}
\newcommand{\python}{\textsc{python}}

\newcommand{\Bjband}{$B_j$-band}

\newcommand{\halpha}{H$\alpha$}
\newcommand{\hbeta}{H$\beta$}
\newcommand{\othree}{$\mathrm{O[III]}\,\lambda\,500.7\,\mathrm{nm}$}
\newcommand{\ntwo}{$\mathrm{N[II]}\,\lambda\,658.3\,\mathrm{nm}$}

\newcommand{\onesigma}{$1\,\sigma$}

\begin{document}

\title{The H IX galaxy survey III: The gas-phase metallicity in \hi\ eXtreme galaxies }

   \author{K. A. Lutz \inst{1,2}
           \and V. Kilborn \inst{2}
           \and B. Catinella\inst{3, 4}
           \and L. Cortese\inst{3, 4}
           \and T. H. Brown\inst{2, 5}
           \and B. Koribalski \inst{6}
            }

\institute{Universit\'e de Strasbourg, CNRS, CDS, Observatoire astronomique de
    Strasbourg, UMR 7550, F-67000 Strasbourg, France
    \email{research@katha-lutz.de}
    \and
    Centre for Astrophysics and Supercomputing, Swinburne University of
    Technology, P.O. Box 218, Hawthorn, VIC 3122, Australia
    \and
    International Centre for Radio Astronomy Research (ICRAR), M468,
    The University of Western Australia, 35 Stirling Highway, Crawley, WA 6009,
    Australia
    \and
    Australian Research Council, Centre of Excellence for All Sky Astrophysics
    in 3 Dimensions (ASTRO 3D), Australia
    \and
    Department of Physics and Astronomy, McMaster University, Hamilton,
        ON L8S 4L8, Canada
    \and Australia Telescope National Facility, CSIRO Astronomy and Space
        Science, P.O. Box 76, Epping, NSW 1710, Australia\\
    }

\date{Received September 15, 1996; accepted March 16, 1997}

\abstract
{This paper presents the analysis of optical integral field spectra for the
\hi\ eXtreme (\hix) galaxy sample. \hix\ galaxies host at least 2.5\,times more
atomic gas (\hi) than expected from their optical $R$-band luminosity. Previous
examination
of their star formation activity and \hi\ kinematics suggested
that these galaxies stabilise their large \hi\ discs (radii up to 94\,kpc)
against star formation due to their higher than average baryonic specific
angular momentum. A comparison to semi-analytic models further showed that
the elevated baryonic specific angular momentum is inherited from the high spin
of the dark matter host.}
{In this paper we now turn to the gas-phase metallicity
as well as stellar and ionised gas kinematics in \hix\ galaxies to gain
insights into recent accretion of metal-poor gas or recent mergers. }
{We compared the stellar, ionised, and atomic gas kinematics, and
examine the variation in the gas-phase metallicity throughout
the stellar disc of \hix\ galaxies }
{We find no indication for counter-rotation in any of the components, the
central metallicities tend to be lower than average, but as low as expected for
galaxies of similar \hi\ mass. Metallicity gradients are comparable to
other less \hi-rich, local star forming galaxies. }
{We conclude that \hix\ galaxies show no conclusive evidence for recent major
accretion or merger events. Their overall lower metallicities are likely due to
being hosted by high spin halos, which slows down their evolution and thus
the enrichment of their interstellar medium. }


\maketitle
\section{Introduction}
\label{sec:intro5}
The study of outliers to scaling relations may inform of the physical
processes that underlay the scaling relation. In the case of the relations
between atomic hydrogen (\hi) and stellar content of galaxies, the study of galaxies that are
more \hi-rich than expected from the stellar or optical properties helps
understand how galaxies acquire and maintain their \hi\ content. Possible
scenarios are more efficient gas accretion or less efficient star formation
than in average galaxies.

Previous surveys of \hi-rich galaxies include \highmass\
\citep{Huang2014}, \bluedisk\ \citep{Wang2013}, \hi\ monsters \citep{Lee2014},
\hi\ excess galaxies \citep{Gereb2016,Gereb2018}, and a sample of \hi-rich,
massive galaxies \citep{Lemonias2014}. While the \highmass, \hi\ monsters, and
\citet{Lemonias2014} samples investigate generally similar galaxies, \bluedisk\
galaxies are systematically less \hi-rich than \hix\ galaxies (see also
\citealp{Lutz2017}). \hi-excess systems differ from other \hi-rich galaxies as
they were selected for being \hi-rich for their specific star formation
rates. For five \highmass\ and all \citet{Lemonias2014} galaxies resolved \hi\
observations were observed, analysed, and published. Out of the five \highmass\
galaxies, three galaxies have a higher than average spin parameter, one
galaxy is on the brink of a star burst, and one galaxy (a group central)
accreted gas from its neighbouring galaxies
\citep{Hallenbeck2014,Hallenbeck2016}. The \hi\ column densities in
the galaxies of \citet{Lemonias2014} are very low and thus star formation is
suppressed. So far the gas-phase metallicity has not been taken into account to
investigate possible gas accretion onto \hi-rich galaxies.

When comparing the current cold gas content (atomic and molecular hydrogen,
i.e. \hi\ and H$_{2}$) of local star forming galaxies to
their star formation activity, \citet{Saintonge2017}, \citet{Schiminovich2010},
and others have found that these galaxies would use their entire cold gas
(\hi\ plus H$_{2}$) content,   on average within 4\,Gyr. Hence, for
galaxies to remain star formers in the future, they need to replenish their
cold gas reservoir. Currently two main avenues of cold gas accretion
are commonly considered: gas accretion from the intergalactic medium
\citep{Birnboim2003,Dekel2006,Keres2005,vandeVoort2011a} or the
infall of satellite galaxies, which bring their own cold gas reservoir with
them \citep{DiTeodoro2014,vandeVoort2011a}. Additionally, the so-called
Galactic Fountain could also increase the cold gas content of a galaxy in the
following way: through stellar feedback gas that is expelled from the disc into the
halo, where hot halo gas condenses onto the expelled parcels of gas and
together they rain back onto the galaxy disc
\citep{Oosterloo2007,Fraternali2007}. Furthermore,
this mechanism redistributes gas within the galaxy and its halo.

The gas-phase metallicity (12 + log(O/H)) is an important indicator of the mode
of accretion. One line of argument that suggests that galaxies
should  accrete metal-poor gas from the intergalactic medium (IGM) is the
comparison of chemical evolution models of galaxies to observations of the
gas-phase metallicity in galaxies. When assuming a closed box model, where the
interstellar medium (ISM) of galaxies is not diluted by pristine gas from the
IGM, metallicities of modelled galaxies are overestimated with respect to
observed metallicities \citep{VandenBergh1962,Kudritzki2015}. Hence, galaxies
need to accrete relatively metal-poor gas from the IGM to dilute their ISM.

An important observational probe into the chemical evolution of galaxies is the
mass--metallicity relation. This relation connects the stellar mass of a galaxy
to the central gas-phase metallicity \citep{Tremonti2004}. It was
shown that the scatter of the mass--metallicity relation is correlated with the
star formation rate of galaxies (SFR) \citep{Mannucci2010} or the \hi\ content
of galaxies \citep{Hughes2013,Bothwell2013,Lagos2016,Brown2018}. Growing
evidence now points to \hi\ being the primary driver and SFR being a
secondary effect due to the dependence of star formation on gas (e.g.
\citealp{Lagos2016}, \citealp{Brown2018}).

N-body, smoothed particle hydrodynamical simulations (e.g.
\citealp{Dave2013}), cosmological simulations (e.g. \citealp{Torrey2019}), and
simple models (e.g. \citealp{Forbes2014a}) have found that the
mass--metallicity relation can be explained if galaxies are seen as systems in
equilibrium between gas accretion, star formation, and outflows. Stochastic
accretion of pristine gas removes galaxies from the equilibrium: the gas
content is increased, the ISM diluted, and star formation triggered. Thus, the
galaxy initially moves from the centre of the mass--metallicity relation to a
more metal-poor position for its stellar mass. With the triggered star
formation, the galaxy now grows in stellar mass and its ISM is gradually
enriched again (through feedback). Thus, it moves back to the equilibrium line
of the mass--metallicity relation.

The observed mass--metallicity relation was originally defined using only the
central metallicities of galaxies. For example the \citet{Tremonti2004}
relation is based on SDSS spectra, which were observed with a
3\,arcsec fibre. Hence, only the metallicity of the central 3\,arcsec is
included in their relation. However, the metallicity of the disc and outskirts
of the galaxy also provide vital information on the chemical evolution.
Recently, a number of large samples of galaxies have been observed with
long-slit or integral field spectra, examples are the SAMI survey
(Sydney-AAO Multi-object Integral field spectrograph galaxy survey;
\citealp{Croom2012,Fogarty2012}), the CALIFA survey (Calar Alto
Legacy Integral Field spectroscopy Area survey; \citealp{Sanchez2012}), the
MaNGA survey (Mapping Nearby Galaxies at APO survey; \citealp{Bundy2015}),
or follow-up of galaxies \citep{Moran2012} in the
\textit{GALEX} Arecibo SDSS survey (GASS; \citealp{Catinella2010,
Catinella2013}). Using these data, a metallicity gradient (i.e. the gas-phase
metallicity change with radius) can be measured. The gas-phase metallicity in
regular spiral galaxies decreases towards the edges of the disc. In the CALIFA
survey, \citet{Sanchez2014} found a universal metallicity gradient in the sense
that the steepness of the gradient is not dependent on the stellar
mass\footnote{The galaxies in this sample cover a stellar mass range
of  $
9.0 \le \log {\rm M}_{\star} [{\rm M}_{\odot}] \le 11.2$, but the majority are
more massive than $10^{10}$ M$_{\odot}$ \citep{Sanchez2013}}. They suggested
that this result indicates a similar chemical evolution in all disc galaxies.
Hence, the metallicity at a given radius is set by the evolutionary  state at that
radius rather than the basic properties of the galaxy. As galaxies form
from the inside  out, the outskirts of a galaxy would be less evolved and
thus more metal-poor. These findings were confirmed by \citet{Ho2015}
and \citet{Kudritzki2015}, who were able to reproduce the uniform
metallicity
gradients with a simple chemical evolution model including in- and outflows.
These models indicate that the metallicity is dependent on the local stellar-to-gas mass ratio, which has also been observed in MaNGA galaxies
\citep{Barrera-Ballesteros2018} (only using estimates of gas mass from dust
attenuation). Other results based on the MaNGA survey, show a
metallicity gradient that varies with stellar mass \citep{Belfiore2017a}. They
suggest that flatter gradients in low mass galaxies show that strong
feedback, gas mixing, and wind recycling must also be important in these
galaxies (i.e. that a mechanism like the Galactic Fountain flattens their
metallicity gradient).

Conversely, \citet{Moran2012} measured the radial metallicity gradient in
massive galaxies from the GASS survey and found the steepest declining
metallicity gradients in galaxies at their lower stellar mass limit of
\logmstar$>10$. They furthermore found that about 10\,\%\ of their
sample have a sharp downturn in metallicity at large radii. This decline is
correlated with the \hi\ content of these galaxies and was interpreted as
a sign for a phase of active gas inflow and disc-building. Similarly,
\citet{SanchezAlmeida2014a} reported some star forming regions in dwarf
galaxies to be more metal-poor than their surroundings. They and
\citet{Ceverino2016} subsequently suggested that these could be
star forming regions that formed in parcels of recently accreted metal-poor
gas.

To date not many samples in addition to the GASS sample and the sample in
this work have actually measured gas properties. In particular, the overlap
between large integral field unit (IFU) surveys (SAMI, MaNGA, and
CALIFA) and \hi\ surveys is still limited for single-dish surveys and close to non-existent
for interferometric (i.e. spatially resolved) \hi\ data.

We  previously compiled a sample of galaxies that contain at least
2.5\,times more \hi\ than expected from their optical $R$-band luminosity.
For these \hi\ eXtreme galaxies (\hix\ galaxies) and a \control\ sample, we
analysed the star formation activity \citep{Lutz2017} and the \hi\ kinematic
properties \citep{Lutz2018}. We found that \hix\ galaxies maintain their \hi\
reservoir by retaining large amounts of \hi\ outside the stellar disc. There
the gas cannot move towards central, denser regions and is stabilised against
star formation due to its high specific angular momentum, which is likely
inherited from a high spin halo. To investigate possible accretion of
metal-poor gas or recent mergers with the \hix\ galaxies, we acquired optical
integral field spectra within the stellar discs of \hix\ galaxies. The data
presented in this paper allow us to compare spatially resolved gas-phase oxygen
abundance measurements to spatially resolved \hi\ observations as well as the
kinematics of different galaxy components.

In addition to  the high angular momentum scenario, another scenario that might
explain the \hi-richness of \hix\ galaxies is very effective or active gas
accretion, which increases the \hi\ content of \hix\ galaxies. This might
either be a temporarily high accretion rate of gas from the IGM or a recent
gas-rich merger. To probe these scenarios, this paper compares \halpha\ and
stellar kinematics to the \hi\ kinematics. Any misalignments in the rotation of
these components can point towards recent mergers \citep{Corsini2014}. We note,
however, that the lack of misalignments cannot exclude recent merger or
accretion events. Furthermore, the central gas-phase metallicity of \hix\
galaxies is examined using the mass--metallicity relation and the steepness of
the gas-phase metallicity gradient in \hix\ galaxies is compared to the
literature. Both measures can also inform of recent inflow of pristine gas.

This paper is structured as follows: in Sect.~\ref{sec:wifes-obs}, details of
observations and the data analysis are described. In Sect.~\ref{sec:results},
results from the analysis of the optical integral field spectra (IFS)
data are presented. These
results are discussed and summarised in Sect.~\ref{sec:discussion_5}.

\section{Galaxy samples and data analysis}
\label{sec:wifes-obs}
\subsection{The HIX galaxy survey: sample selection}

\begin{table*}
    \centering
    \begin{tabular}[H]{l || c c c c c c c c c c }
ID & RA & D & log M$_\star$ & log M$_{HI}$ & R$_{HI}$ & R$_{25}$ & V$_{sys}$ &
W50 & V$_{rot}$ & WiFeS obs?\\
~ & Dec & ~ & ~ & ~ & ~ & ~ & ~ & ~ & ~ & ~ \\
~ & deg & Mpc & M$_\odot$ &  M$_\odot$ & kpc & kpc & \kms & \kms & \kms & ~ \\
(1) & (2) & (3) & (4) & (5) & (6) & (7) & (8) & (9) & (10) & (11)\\
\hline
ESO111-G014 & 2.0782 & 112 & 10.5 & 10.7 & 49.6 & 23.5 & 7784 & 353 & 197 &
Yes  \\
~ & -59.5156 & ~ & ~ & ~ & ~ & ~ & ~ & ~ & ~ & ~ \\
ESO243-G002 & 12.3938 & 129 & 10.7 & 10.5 & 55.6 & 16.4 & 8885 & 284 & 167 & No
\\
~ & -46.8744 & ~ & ~ & ~ & ~ & ~ & ~ & ~ & ~ & ~ \\
NGC\,289 & 13.1765 & 23 & 10.5 & 10.3 & 86.9 & 11.0 & 1627 & 277 & 168 & Yes\\
~ & -31.2058 & ~ & ~ & ~ & ~ & ~ & ~ & ~ & ~ & ~ \\
ESO245-G010 & 29.1853 & 82 & 10.5 & 10.4 & 50.9 & 23.5 & 5752 & 376 & 190& Yes
\\
~ & -43.9725 & ~ & ~ & ~ & ~ & ~ & ~ & ~ & ~ & ~ \\
ESO417-G018 & 46.8050 & 67 & 10.3 & 10.4 & 45.5 & 19.9 & 4745 & 327 & 174 &
Yes\\
~ & -31.4007 & ~ & ~ & ~ & ~ & ~ & ~ & ~ & ~ & ~ \\
ESO055-G013 & 62.9290 & 105 & 10.2 & 10.4 & 41.0 & 8.1 & 7387 & 216 & 199 &
Yes\\
~ & -70.2331 & ~ & ~ & ~ & ~ & ~ & ~ & ~ & ~ & ~ \\
ESO208-G026 & 113.8380 & 40 & 9.8 & 9.8 & 28.8 & 6.6 & 2979 & 277 & 132 & Yes\\
~ & -50.0430 & ~ & ~ & ~ & ~ & ~ & ~ & ~ & ~ & ~ \\
ESO378-G003 & 172.0167 & 41 & 10.1 & 10.2 & 44.3 & 10.3 & 3022 & 260 & 126 &
Yes\\
~ & -36.5427 & ~ & ~ & ~ & ~ & ~ & ~ & ~ & ~ & ~ \\
ESO381-G005 & 190.1363 & 80 & 10.1 & 10.2 & 37.8 & 11.7 & 5693 & 216 & 114 &
Yes\\
~ & -36.9681 & ~ & ~ & ~ & ~ & ~ & ~ & ~ & ~ & ~ \\
ESO461-G010 & 298.5182 & 98 & 10.1 & 10.4 & 23.7 & 17.3 & 6701 & 345 & 154 & No
\\
~ & -30.4843 & ~ & ~ & ~ & ~ & ~ & ~ & ~ & ~ & ~ \\
ESO075-G006 & 320.8729 & 154 & 10.6 & 10.8 & 94.0 & 21.0 & 10613 & 304 & 217 &
Yes\\
~ & -69.6848 & ~ & ~ & ~ & ~ & ~ & ~ & ~ & ~ & ~ \\
ESO290-G035 & 345.3853 & 84 & 10.5 & 10.3 & 36.3 & 23.7 & 5882 & 383 & 179 &
Yes\\
~ & -46.6463 & ~ & ~ & ~ & ~ & ~ & ~ & ~ & ~ & ~ \\ \hline
IC\,4857 & 292.163239   & 67 & 10.5 & 10.0 & 29.2 & 17.1 & 4669 & 282 & 156 &
Yes \\
~ & -58.767879 & ~ & ~ & ~ & ~ & ~ & ~ & ~ & ~ & ~ \\
        \end{tabular}
\caption[The HIX galaxies]{Basic properties of the \hix\ and \control\
(below the line) galaxies:
\textbf{Column (1)}: Identifier for the galaxy. \textbf{Column (2)}: Right
ascension and Declination in degrees for J2000 epoch from the 2MASX catalogue
\citep{Skrutskie2006}.
\textbf{Column (3)}: Luminosity distance in Mpc based on the \hipass\ systemic
velocity and above mentioned cosmological parameters. \textbf{Column (4)}:
Stellar mass based on the 2MASX $K_s$-band flux \citep{Skrutskie2006} and the
\citet{Wen2013} prescription. \textbf{Column (5)}: \hi\ mass based on flux
measurements from
the ATCA data cubes. \textbf{Column (6)}: Radius of \hi\ disc at 1\msunpcsq\
isophote in kpc. \textbf{Column (7)}: Radius of stellar disc at 25\magasec\
isophote in kpc \citep{Lauberts1989}. \textbf{Column (8)}: Systemic velocity in
\kms\ from tilted ring fit. \textbf{Column (9)}: 50\,\%\ width in \kms\
from \hipass. \textbf{Column (10)}: Rotation velocity from tilted ring fit.
\textbf{Column (11)}: Whether (Yes) or not (No) there are any observations with
the WiFeS spectrograph for this galaxy.  The data in Cols. (5), (6), (8), and (10) come from the analysis in
\citet{Lutz2018}. }
\label{tab:hix_prop}
\end{table*}

The sample selection was described in detail in \citet{Lutz2017}; here we
give a brief summary: \hix\ galaxies are selected from the \hi\ Parkes All Sky
Survey (\hipass, \citealp{Barnes2001}). In particular, we used a high quality,
high fidelity catalogue of galaxies with both \hi\ mass measurement and optical
photometry \citep{Denes2014}, which was based on the \hipass\ catalogues
\citep{Barnes2001,Meyer2004,Zwaan2004,Koribalski2004} and the optical
counterparts in the \hopcat\ catalogue \citep{Doyle2005}.

Based on the \citet{Denes2014} scaling relation between \hi\ mass and $R$-band
luminosity, \hix\ galaxies were selected to fulfil the following criteria:
\begin{itemize}
    \item they host at least 2.5\,times more \hi\ than expected from their $R$-band
    luminosity, i.e. they lie at least $1.4\,\sigma$ above the scaling relation;
    \item they are located south of Dec < -30\,deg for good observability with the
    Australian Telescope Compact Array;
    \item they are brighter in absolute $K_s$-band magnitude than -22\,mag to
    restrict the sample to massive spiral galaxies.
\end{itemize}

For comparison, a \control\ sample was compiled from the same catalogue with the
same selection criteria except that galaxies should have between 1.6\,times
less and more \hi\ than expected from the \citet{Denes2014} $R$-band scaling
relation (i.e. within $\pm0.7\sigma$).

An overview of basic properties of \hix\ and \control\ galaxies
examined in this paper is given in Table~\ref{tab:hix_prop}.

\subsection{Observations of HIX galaxies with WiFeS}
We used the WIde FiEld Spectrograph (\wifes, \citealp{Dopita2007}) on the ANU
2.3\,m telescope in Siding Spring, Australia to obtain optical integral
field spectra of 10 out of 12 HIX galaxies and one \control\ galaxy.

\begin{figure*}
    \centering
    \includegraphics[width=6.2in]{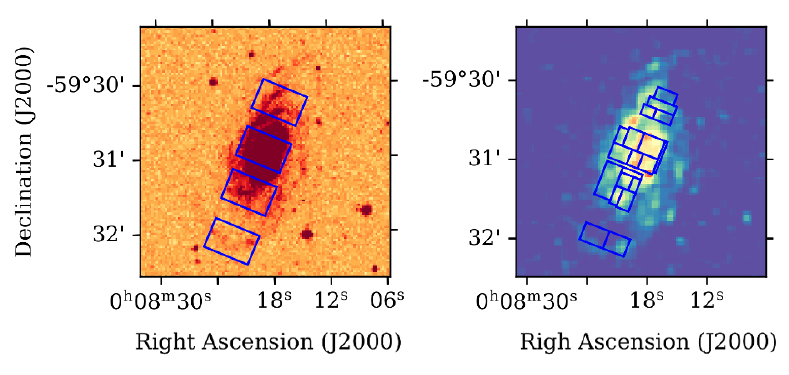}
    \caption[Example of \wifes\ data]{Example of \wifes\ data for ESO111-G014.
        \textbf{Left:} Red-scale image in the background shows the \Bjband\
        optical image. The overlaid blue squares are of the size of the \wifes\
        aperture and are located  where the telescope was pointed.
        \textbf{Right:} Colour map in the background shows the near-UV image,
        the overlaid blue squares indicate where `single' star forming
        regions were defined. (Similar figures for all galaxies can be
        found in Appendix~\ref{app:op_spec})}
    \label{fig:eso111-wifes}
\end{figure*}

\wifes\ is an image slicing integral field unit, meaning that it consists
of 25\,slitlets, each $1\times36$\,arcsec in size. Combining the length of the
slitlets and their number, \wifes\ has a $36\times25$\,arcsec$^{2}$ field of
view. This is smaller than the optical disc of the \hix\ sample galaxies, which
have an average 25\magasec\ isophote radius of 47\,arcsec. Therefore, the aim
was to observe every galaxy with multiple pointings to cover the
centre and the outer regions of the stellar disc.
Figure~\ref{fig:eso111-wifes} shows an example of multiple pointings
towards ESO111-G014. In the left panel the \wifes\ pointings are shown,
while the right panel details the final binning of the pointings.

The aim was to to obtain 60\,min (90\,min) of on-source time per pointing
for galaxies with average surface brightness brighter (fainter) than
22\magasec. The actual total on-source exposure times of each pointing are
between 15 and 90\,min depending on the signal strength. The
total on-source time per pointing is split into single exposures of
15\,min. Stacking multiple short exposures helps to remove cosmic rays and
small image errors. To subtract the night sky foreground, all science
spectra are taken in the so-called nod-and-shuffle mode, in which the telescope
nods between the science target (i.e. the galaxy) and an `empty' part of the
sky. This way a spectrum of the sky and a sum of the sky spectrum and the
galaxy is obtained. The separate sky spectrum is then subtracted from the
intermingled sky and galaxy spectrum to obtain a pure galaxy spectrum.

Incoming spectra were split by a dichroic at 560\,nm into a red and a
blue half. Thus, a wide wavelength range can be covered in one shot.
\wifes\ was set up with $R=3000$ gratings in  the red and
the blue arm for all observations. At a wavelength of 660\,nm this is equivalent
to a wavelength step of 0.22\,nm per pixel or a velocity step of 100\kms.

Every night of observations was completed with standard calibration
images including bias, sky flat field, dome flat field, wire imaging
for centring the slitlets, and NeAr and CuAr arc lamp spectra for wavelength
calibration. Two to three times a night a standard star was observed
for flux calibration. Dark current information was obtained from
overscan regions (i.e. parts of the CCD that were not exposed). The
observations were conducted between August 2014 and April 2016 under
photometric conditions. Details about
the number of pointings and on-source exposure times are given in
Table~\ref{tab:hix_wifes_obs}.

\begin{table*}
    \centering
        \begin{tabular}[H]{l || c c c c c c }
            ID       & Central & South/ East & North/ West & Arm/ Dwarf &
            Outer South & Outer North \\
            (1) & (2) & (3) & (4) & (5) & (6) & (7)  \\  \hline \hline
            ESO111-G014& 60  & 60  & 60  & 45 & --- & --- \\
            NGC\,289   & 15  & 45  & 45  & --- & 45  & 60 \\
            ESO245-G010& 60  & --- & --- & --- & --- & --- \\
            ESO417-G018& --- & 75  & --- & --- & --- & --- \\
            ESO055-G013& 30  & --- & --- & --- & --- & --- \\
            ESO208-G026& --- & 75  & 90  & --- & --- & --- \\
            ESO378-G003& 45  & 60  & 60  & --- & --- & --- \\
            ESO381-G005& 30  & 45  & 60  & 60  & --- & --- \\
            ESO075-G006& 30  & 90  & 60  & --- & --- & --- \\
            ESO290-G035& 30  & 60  & 60  & --- & --- & --- \\  \hline
            IC\,4857   & 90  & 15  & 60  & --- & --- & --- \\

        \end{tabular}
        \caption[The \wifes\ observations]{The \wifes\
            observations of the \hix\ sample and \control\ galaxy IC\,4857.
            \textit{Column (1)} gives the ID of the galaxies. \textit{Columns
            (2)} to \textit{(7)} give the exposure times in minutes for single
            pointings. The name of the column describes where the pointing is
            located with respect to the galaxy centre. For a visualisation of
            pointing location see Fig.~\ref{fig:eso111-wifes} for ESO111-G014
            and Appendix~\ref{app:op_spec} for all other galaxies.}
        \label{tab:hix_wifes_obs}
\end{table*}

\subsection{Data reduction of WiFeS observations}
\citet{Childress2014} provide the fully automated \pywifes\ pipeline for
\wifes\ data. This pipeline includes bad pixel repair, bias and dark current
subtraction, flat fielding, wavelength calibration, sky subtraction, flux
calibration, and data cube creation. The resulting data cubes are 70 by
25 pixels (35 by 25\,arcsec) in size and their wavelength range is set
from 650.0 to 685.0\,nm and 460.0 to 525.0\,nm for the red and blue cubes,
respectively. We note  that these measures imply a pixel size of 1\,arcsec in one
spatial direction and 0.5\,arcsec in the other spatial direction.

\begin{figure*}
    \centering
    \includegraphics[width=6.2in]{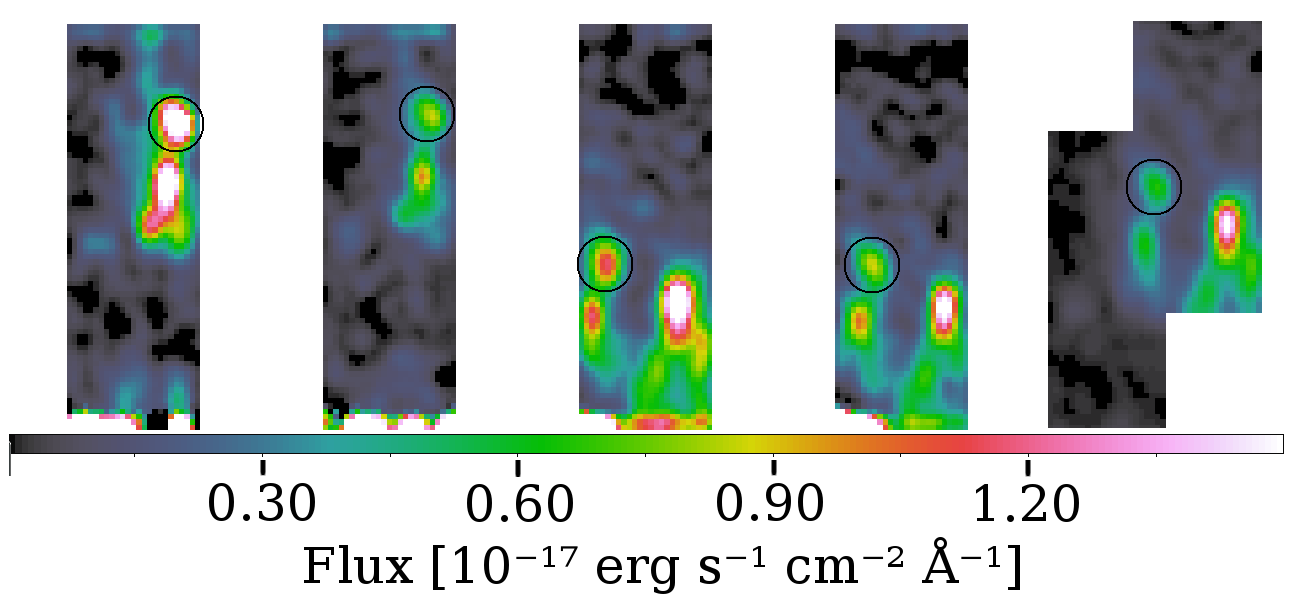}
    \caption[Example of stacking \wifes\ data cubes]{Colour-scale images in
        the four left panels show the plane with \halpha\ emission in the four
        single data cubes of the southern pointing of ESO111-G014. The black
        circles indicate the feature that was used to align the four data cubes.
        The right panel shows the plane with \halpha\ emission in the final,
        stacked data cube.}
    \label{fig:ngc289-wifes-stack}
\end{figure*}

Single galaxy exposures are spatially offset from each other. To account for
this, \pywifes\ was run on each observed galaxy exposure individually.
For each pointing, data cubes were then median stacked after the full
data reduction by \pywifes. To align single data cubes of one pointing, the
distributions of \halpha\ emission in the data cubes were compared.
For an example see Fig.~\ref{fig:ngc289-wifes-stack}. The locations of bright
centres of star forming regions were matched up  between the
different exposures of one pointing. Once the pixel offset between the single
data cubes per pointing was determined, the cubes were median
stacked.

After stacking, cubes were reshaped such that both spatial dimensions
have pixel sizes of 1\,arcsec. For this task the \textsc{scipy}
\citep{Jones2001} function \texttt{scipy.interpolate.griddata} was
used. Data were linearly interpolated to the new pixel size of
$1\times1$\,arcsec.

Initially, observations of this \wifes\ project do not include any
world coordinate system (WCS) information in their file headers. To
add this information in the file headers the following procedure
was applied: right before the \wifes\
observations were taken, an image with the acquisition and guiding (A\&G)
camera was taken. In the centre of this image the shadow of the \wifes\ IFU is
visible. We applied an astrometric solution to this A\&G camera image
using bright stars and the Aladin software \citep{Bonnarel2000}. Since
  the pixel size of the IFU cube (1\,arcsec) and the location of the IFU
within the A\&G camera image (due to its shadow) are both known, we then simply
transferred the astrometric solution from the A\&G camera image to the IFU
cubes. After this procedure, we measured the position of prominent
features (e.g. peaks of star formation, stars) in collapsed maps of the WiFeS
cubes and compared them to the positions of these features measured on the
SuperCOSMOS $B_j$-band images. We find an average distance between the two
position measurements of ($3\pm6$)\,arcsec.

This procedure resulted in two data cubes for each pointing: one
containing the red half of the spectrum and the other   the blue half. To
increase the signal-to-noise ratio of spectra, the spaxels in both cubes
were binned such that one bin encompasses a star forming region and
 the \halpha\ and \ntwo\ emission lines are visible in the resulting
spectrum. These star forming regions were defined by hand on the
\halpha\ emission planes of the \wifes\ data cubes to be a connected region of
\halpha\ emission (see Fig.~\ref{fig:eso111-wifes}).

In pointings towards the edges of stellar discs, the one-dimensional
spectra of individual star forming regions do not pick up emission from
the stellar continuum. In those cases the background was estimated as a
constant from emission and sky line free parts of the spectrum
\citep{Wisnioski2015}. For our data set we chose the following wavelength
ranges: Indicating with $\lambda_\alpha$ and $\lambda_{N}$ the redshifted
wavelengths of \halpha\ and \ntwo\ lines, respectively, we chose the red window
as [$\lambda_N$ + 2\,nm, $\lambda_N$ + 7\,nm] if $\lambda_\alpha < 665$\,nm,
and [$\lambda_\alpha$ - 8\,nm, $\lambda_\alpha$ - 3\,nm] otherwise. The blue
window is always [$\lambda_\beta$ - 8\,nm, $\lambda_\beta$ - 3\,nm], where
$\lambda_\beta$ is the redshifted wavelength of the \hbeta\ line. We then took
all the available data within these windows and calculated the
40th and 60th percentiles. The mean of the data in the 40th to 60th percentile range was
then set as the constant background value for those spectra that do not show
any stellar continuum.

If stellar continuum is detected in a spectrum of a star
forming region, then the full spectrum (i.e. the blue and the red half)
was used to fit a stellar population synthesis model and gas emission
lines with the penalized pixel-fitting (pPXF) method and the \python\
script by \citet{Cappellari2004} and \citet{Cappellari2017}. pPXF used the stellar
population synthesis models from the \citet{Vazdekis2010} library plus Gaussian
emission lines. From the location of stellar absorption features, the stellar
recession velocity in this star forming region was measured.
pPXF provides the line strength for each modelled emission
line (\halpha, \hbeta, \othree,\ and \ntwo). For the recession velocity of the
gas it was assumed that there is only one kinematic component, and the recession
velocity was modelled from all four emission lines simultaneously (i.e. one
recession velocity measurement from the four emission lines).

The emission line fluxes were corrected for internal dust attenuation
assuming the \cite{Calzetti2000} attenuation curve and $\rm R'_V=4.05$ (see
also \citealp{Moran2012}). From the corrected \halpha, \hbeta, \othree, and
\ntwo\ emission line fluxes, the gas-phase oxygen abundances were
determined with the $O3N2$ and $N2$ method as described by \citet{Pettini2004}.

We tested several options and found for spectra, in which the
stellar continuum could be modelled with pPXF, that metallicity measurements
based on the subtraction of a constant background and the $N2$ method
agree with metallicities based on subtraction of a pPXF modelled stellar
continuum and the $O3N2$ method. Hence, wherever possible we use $O3N2$
metallicities calculated from pPXF modelled emission line fluxes, otherwise the
$N2$ method with the emission lines measured after subtraction of a
constant background.

To de-project the on-sky galactocentric radius $r$, first the angle
$\theta$ between the galaxy semi-major axis and the line between galaxy centre
and star forming region was determined. The de-projected radius
$r_{dpj}$ was then set to
\begin{equation}
    r_{\rm dpj} = \frac{r \times \sqrt{(b \cos \theta)^2 + (a \sin \theta)^2}}
    {b},
\end{equation}
where $a$ and $b$ are the semi-major and semi-minor axis, respectively.
Metallicity gradients were measured by fitting a line to the
metallicities as a function of de-projected galactocentric radius. The slope of
this line is the metallicity gradient.

\begin{table*}
    \centering
    \begin{tabular}[H]{p{1.5cm} || p{1.5cm} | p{3.0cm} | p{1.0cm} | p{1.0cm} |
    p{1.5cm} | p{1.0cm} | p{1.0cm} | p{3cm} }
        Sample & Number of galaxies used here & relevant references & optical
        spectra & \hi\ masses & metallicity indicator & min log M$_{\star}$
        [M$_\odot$] & max log M$_\star$ [M$_\odot$]  & Notes  \\
        (1) & (2) & (3) & (4) & (5) & (6) & (7) & (8) & (9) \\
        \hline \hline
    \hix\ & 10 & \citet{Lutz2017, Lutz2018} & Y & Y & PP04 O3N2, N2 & 9.8 &
    10.6 &  \\ \hline
    IC\,4857 & 1 & as \hix\  & Y & Y & PP04 O3N2, N2 & 10.5 & & member of the
    original \control\ sample \\ \hline
    \hipass\ & 1796 &  \citet{Denes2014,Barnes2001,Meyer2004,Doyle2005}
    & N & Y & - & 8.0 & 11.5 & biased towards \hi-rich objects, maximal
    redshift ~0.04, parent sample for \hix\ galaxies \\ \hline
    SDSS MPA-JHU & 60021 & \citet{Abazajian2009,Salim2007} & Y & N & PP04 O3N2
        & 8.0 & 11.5 & to provide a benchmark mass-metallicity relation  \\
        \hline
    GASS  with longslit spectroscopy & 96  & \citet{Catinella2018,Moran2012} &
    Y & Y & PP04 O3N2 & 10.0 & 10.82 & longslit spectra typically out to 1
    Petrosian 90\,\%\ radius, see text for sample selection\\
        \hline
    CALIFA & 227 & \citet{Sanchez2014} & Y & N & PP04 O3N2 & 9.0 & 11.2 &
        IFU data, the metallicity gradient is typically measured between 0.5
        and 2 effective radii.
    \end{tabular}
    \caption{Summary of the samples and data used in this paper.
    \textbf{Column (1)}: name of the sample. \textbf{Column (2)}: number of
    galaxies used in
    this paper. \textbf{Column (3)}: references to the papers in which the
    sample and/or  data
    are described in more detail. \textbf{Column (4)}: whether (Y) or not (N)
    optical
    spectra are available for this sample. \textbf{Column (5)}: whether (Y) or
    not (N)
    measurements of the \hi\ mass are available for this sample.
    \textbf{Column(6)}:
    which metallicity indicator was used to estimate the gas-phase
    metallicity. PP04 refers to \citet{Pettini2004}. \textbf{Columns (7) and
    (8)}: the
    minimum and maximum stellar mass limit of the sample, respectively.
    \textbf{Column
    (9)}: Further notes.}
\end{table*}

\begin{figure}
    \centering
    \includegraphics[ width=3.15in]
    {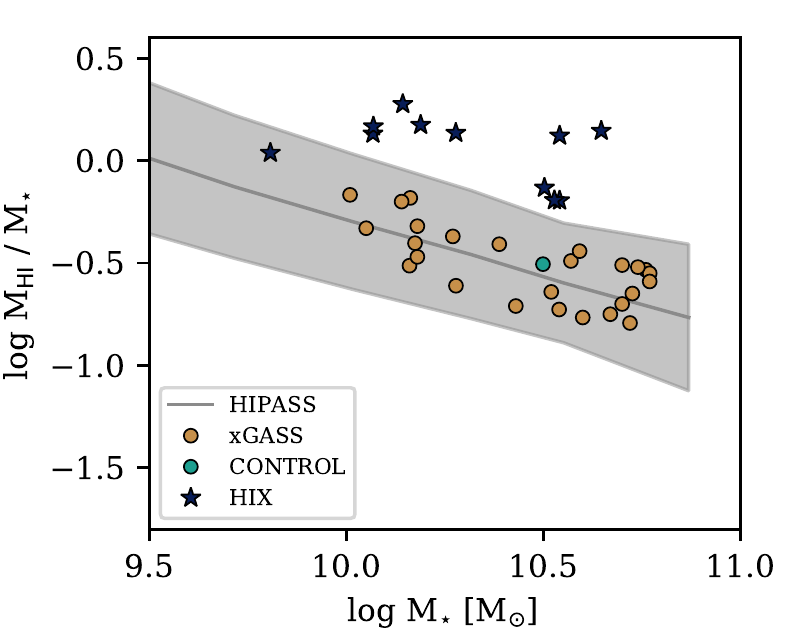}
    \caption[The \hi\ gas mass fraction as a function of stellar mass]{The \hi\
    gas mass fraction as a function of stellar mass for \hix\ galaxies (blue
    stars), \control\ galaxy IC\,4857 (turquois circle), and those galaxies
    from GASS with an
    optical long-slit spectrum (yellow dots). The grey line is the running
    average of the \hipass\ parent sample and the grey shaded area the
    corresponding \onesigma\ scatter. }
    \label{fig:fhi_vs_mstar}
\end{figure}

\subsection{Comparison samples}
\label{sec:comp_samples}
As only one of the original \control\ galaxies was observed with
\wifes, data from the Sloan Digital Sky Survey (SDSS, \citealp{York2000})
data release 7 \citep{Abazajian2009}, and the (extended)
GASS \citep{Catinella2018,Catinella2013,Catinella2010,Moran2012} are used for
comparison.

For the SDSS data, line emission and stellar mass measurements
\citep{Kauffmann2003a,Salim2007} from the MPA-JHU
SDSS DR7\footnote{\url{https://wwwmpa.mpa-garching.mpg.de/SDSS/DR7/}} catalogue
are used. For galaxies within a redshift range of $0.002<z<0.06$ central
metallicities were calculated from $5\,\sigma$ detected emission lines
using again the $O3N2$ method. These data are used as a benchmark
mass--metallicity relation in Sect.~\ref{sec:mass-metal}.

We also include data from the optical follow-up of GASS galaxies.
These data are based on long-slit spectra rather than IFU spectra, and they
cover radii to 1 Petrosian 90\,\%\ radius. The GASS survey
measured the integrated atomic gas content of $\sim 1200$, stellar-mass
selected galaxies. In addition, for about 200 galaxies optical
long-slit spectra were obtained. From these spectra metallicity gradients
were measured  \citep{Moran2012}. Below, metallicity
gradients,
stellar masses, and \hi\ mass fractions of star forming galaxies (i.e. near-UV-r
colour bluer than 4.3) are used for comparison. To mimic the previous
control sample, we restricted the sample to galaxies which are within
$\pm0.7\,\sigma$ of the \hipass\ running average in the \hi\ mass fraction versus
stellar mass plane (see Fig.~\ref{fig:fhi_vs_mstar}). These GASS galaxies are
shown as yellow dots in Fig.~\ref{fig:fhi_vs_mstar}).

\section{Results}
\label{sec:results}
\subsection{Comparison of stellar, H$\alpha$ and H\,{\small I} velocities}
\label{sec:velos}
\begin{figure*}
    \centering
    \includegraphics[ width=6.3in]
    {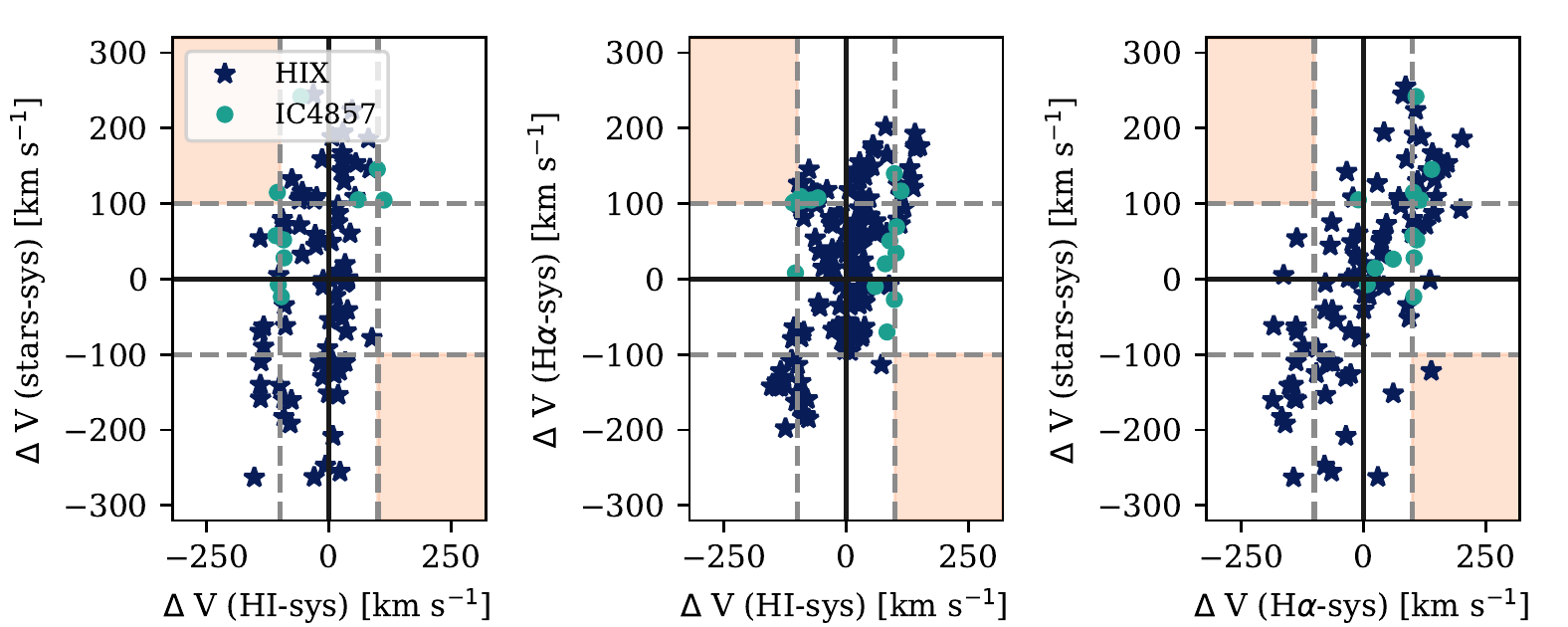}
    \caption[Comparison of velocities]{Comparison of the recession
    velocities of different galaxy components: stars, \hi,\ and ionised gas.
    Every panel compares two components. To account for the different redshifts
    of the different galaxies, the systemic velocity of the respective galaxy
    (see Table~\ref{tab:hix_prop}) was subtracted from the recession velocity
    measurement. No clear sign of counter rotation is visible.  }
    \label{fig:differences_velocities}
\end{figure*}
To compare the kinematics of stars, and ionised and atomic gas, we consider
their recession velocities minus the systemic velocity measured from the titled
ring models in \citet{Lutz2018}. A pairwise comparison of the three components
is shown in Fig.~\ref{fig:differences_velocities}. This figure shows for every
spaxel how the local velocity of \hi\ compares to that of the stars (left
panel), to that of the \halpha\ line (middle panel), and the velocities of
stars to the \halpha\ line (right panel). The horizontal and vertical lines
indicate $\pm100$\kms\ of the systemic velocity, which is the velocity resolution
of the used setup of the \wifes\ spectrograph. Spaxels that are located in the
upper left or lower right corner of this figure (red-shaded area) would
indicate approaching movement in one component and receding movement in the
other and thus counter-rotation. As can been seen, hardly any spaxels are located
in the red regions. If the sign of the velocities of two components is
different, this is usually within 100\kms, which is likely due to the low
velocity resolution of the optical observations. Detailed kinematic maps for
every galaxy can be found in Appendix~\ref{app:op_spec}.

\subsection{{\small HIX} galaxies on the mass--metallicity relation}
\label{sec:mass-metal}
\begin{figure}
    \centering
    \includegraphics[ width=3.15in]
    {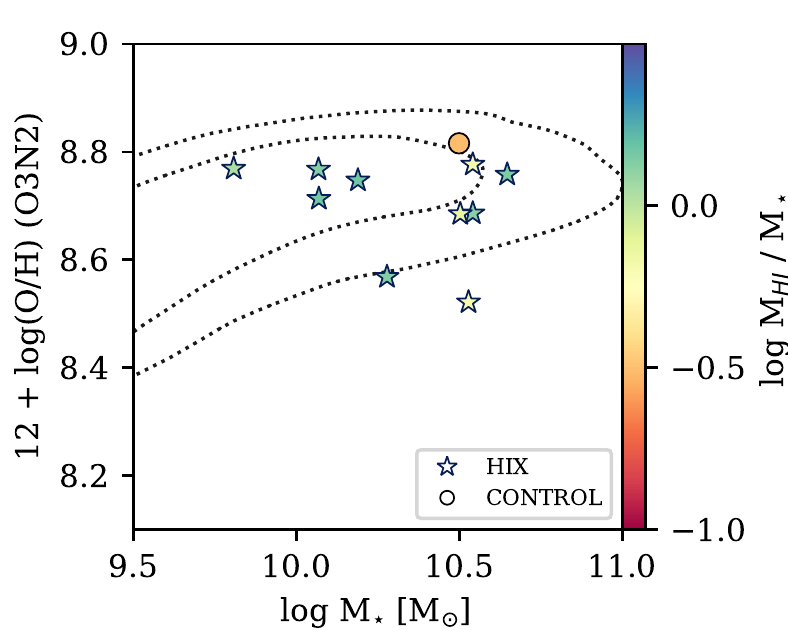}
    \caption[The mass metallicity relation for the \hix\ galaxies]{Mass--metallicity relation for the \hix\ galaxies: we show the gas-phase
metallicity (determined with the O3N2 parameter following \citealt{Pettini2004})
as a function of stellar mass. The contours enclose 64\%\ and 90\,\%\ of
SDSS galaxies. The circle indicates the \control\ galaxy
IC\,4857 and \hix\ galaxies are marked with stars. All symbols are
also colour-coded by the \hi\ mass fraction. }
    \label{fig:metal_vs_mstar}
\end{figure}

Figure~\ref{fig:metal_vs_mstar} shows the mass--metallicity relation for the
\hix\ and \control\ galaxies that were observed with \wifes. As
mentioned above, the scatter of the mass--metallicity relation is driven by the
\hi\ content of the galaxies, where
more \hi-rich galaxies are more metal-poor at a given stellar mass
\citep{Bothwell2013,Brown2018}. In Fig.~\ref{fig:metal_vs_mstar}, the black
contours enclose 64\%\ and 90\,\%\ of the full SDSS DR7 sample (as described
in Sec.~\ref{sec:comp_samples}). The data points showing the \hix\
galaxies (stars) and the \control\ galaxy (large circle) are
colour-coded according to \hi\ mass fraction. For these galaxies, we only
consider the most central metallicity measurement, which does not show
line-ratios that are consistent with active galactic nuclei (AGN) emission.

There are no central \wifes\ pointing (closer than $\approx11$\,kpc to the
centre) available for ESO417-G018. Due to a metallicity gradient, ESO417-G018
is therefore located below the SDSS 90\,\%\ contour. The \hix\ galaxies
with the lowest central metallicity is, however, ESO290-G035. Apart from these
two galaxies and ESO208-G026 (\hix\ galaxy with lowest stellar mass),
\hix\ galaxies are generally located in the lower half of the SDSS scatter
(black dotted lines). To sum it up, \hix\ galaxies are generally expected to
have lower metallicities in the centres, and indeed we find them to be
predominantly located in the low metallicity half of the SDSS scatter.

\subsection{Gas-phase metallicity distribution}
\label{sec:metal-dist}
\begin{figure}
    \centering
    \includegraphics[ width=3.15in]
    {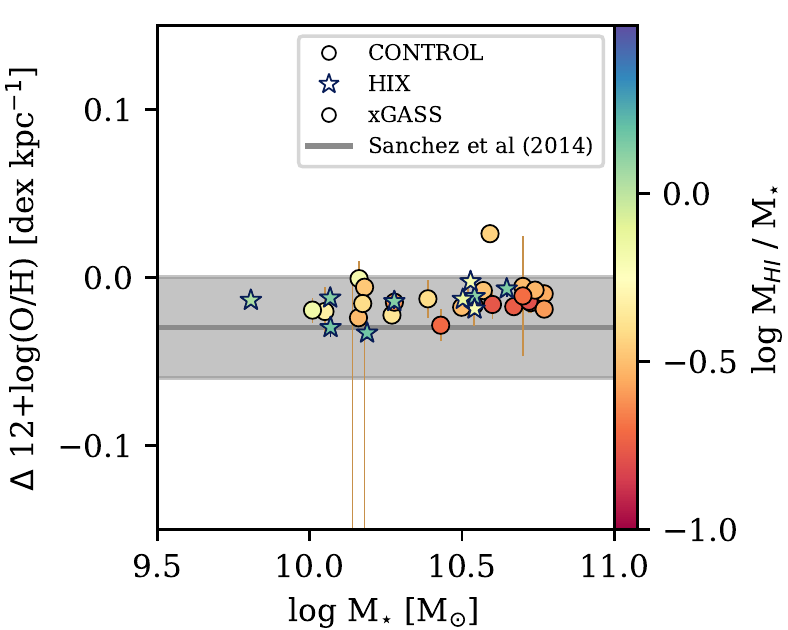}
    \caption[Gas-phase metallicity gradient as a function of stellar mass]{Metallicity gradient in dex\,kpc$^{-1}$ as a function of stellar mass. The grey
line indicates the average gradient by \citet{Sanchez2014} and the grey shaded
area the \onesigma\ scatter. Stars indicate the metallicity gradient measured
in \hix\ galaxies, and the circles the gradients of the control galaxies (IC\,4857
and GASS). All symbols are colour-coded by the \hi\ mass
fraction of the galaxy. Overall, \hix\ and control galaxies behave similarly
to the galaxies in the \citet{Sanchez2014} sample.}
    \label{fig:metal_grad_hist}
\end{figure}

Exploiting the spatial information from the IFU spectral data, we
extracted radial metallicity profiles and measured metallicity
gradients. For a consistent comparison between our samples and the literature,
the metallicity gradients are measured in units of dex\,kpc$^{-1}$. Metallicity
gradients as a function of stellar mass are shown in
Fig.~\ref{fig:metal_grad_hist} and data points are colour-coded by \hi\ mass
fraction. The grey line is the average \citet{Sanchez2014} gradient based on
CALIFA data and the grey shaded area their \onesigma\ scatter, within which all
\hix\ galaxies are located. Hence, \hix\ galaxies do not show steeper
metallicity profiles than average local spiral galaxies. In fact, they tend to
be located at the flat end of the \citet{Sanchez2014} scatter and their
average metallicity gradient is ($-0.016 \pm 0.009$)\,dex\,kpc$^{-1}$. This
means that the average \citet{Sanchez2014} gradient does not agree with the
average \hix\ gradient and its 1\,$\sigma$ scatter, but they do agree within
both their errors.

More recent work on metallicity gradients with SAMI and MaNGA galaxies
generally agrees with \citet{Sanchez2014}, despite detecting
anti-correlations between metallicity gradients and stellar masses at low
stellar masses (log\,M$_\star$ [M$_\odot$] < 9.6)
\citep{Poetrodjojo2018,Belfiore2017a}. We note,
however, that diffuse emission significantly affects metallicity measurements
in IFS data with poor spectral resolution, such as SAMI and MaNGA
\citep{Poetrodjojo2019}. Furthermore, these surveys preferentially (only)
report quantitative gradients in units of dex per effective radius or per
isophotal radius. We thus refrain from more detailed comparison.

The \control\ galaxy IC\,4857 and the galaxies from the GASS
\citet{Moran2012} sample are also mostly located within the \onesigma\
scatter of CALIFA (circles in Fig.~\ref{fig:metal_grad_hist}). We do not see
the variation with stellar mass that was reported by \citet{Moran2012}, perhaps because we only look at a small subsample. For the GASS
galaxies we find an average metallicity gradient of ($-0.01 \pm
0.01$)\,dex\,kpc$^{-1}$, which is consistent with the the average \hix\
gradient, but again smaller than the average CALIFA gradient.

\citet{Moran2012} furthermore suggested that large, abrupt drops (of
about 0.25\,dex) in metallicity at the outskirts of the optical disc in massive
galaxies can be caused by the inflow and distribution of pristine gas from the
outskirts throughout the entire galaxy disc. Similarly,
\citet{SanchezAlmeida2014a} found in a sample of dwarf galaxies, star forming
regions with very low metallicities compared to other star forming regions
within the same dwarf galaxy. They and \citet{Ceverino2016} argue that these
metal-poor star forming regions are induced by pristine gas accretion. There
are neither metallicity drops at the outskirts nor particularly metal-poor star
forming regions observed in the \hix\ and galaxies or in IC\,4857 (for
the detailed data see Appendix~\ref{app:op_spec}).

\section{Discussion and conclusion: optical spectra of {\small HIX} galaxies}
\label{sec:discussion_5}
We have presented the analysis of optical spectra of ten \hix\
galaxies as well as a set of control galaxies. In a first step, measurements of
the \halpha, stellar (where available) and \hi\ recession velocities in the
\hix\ galaxies and IC\,4857 have been compared. Generally, the
kinematics of all three components are similar. While the detection of
counter-rotation, in particular between the stellar and gaseous components,
would be strong evidence of recent accretion of large amounts of gas
\citep{Corsini2014}, the opposite (no counter-rotation means no accretion) is
not necessarily true. Hence, from the kinematics point of view, the evidence
for recent accretion of large amounts of gas in \hix\ galaxies is still
inconclusive.

Using strong emission lines, gas-phase metallicities have been measured
throughout the detected stellar discs of \hix, \control, and
GASS galaxies. These measurements have been used to place these galaxies
on the mass--metallicity relation and to compare their metallicity gradients,
amongst each other and to the average of \citet{Sanchez2014}.

On the mass--metallicity relation, most \hix\ galaxies are located below the
ridge of the relation. This is expected from and consistent with
previous studies that  found that \hi-rich
galaxies are generally metal poor (e.g.
\citealt{Bothwell2013,Hughes2013,Lagos2016,Brown2018}). These studies, based
on observations and on simulations, suggest that galaxies accrete gas
stochastically, and that galaxies below the ridge of the mass--metallicity
relation have recently accreted gas, which increased their gas content and
decreased their gas-phase metallicity.

The radial distribution of the gas-phase metallicity (within the stellar disc)
of \hix\ galaxies appears similar to the one of galaxies with average \hi\
contents. All measured gradients in \hix\ galaxies are negative, i.e. they
are more metal poor in the outskirts than in the centres, which is in
agreement with the scenario where no major merger occurred recently. We have
furthermore compared the \hix\ metallicity gradients to the metallicity
gradients of galaxies with less massive \hi\ discs (drawn from the \control\
sample and GASS). Regardless of their overall \hi\ content,
there is no significant difference in the metallicity gradients of these
galaxies. We note again that all \hix\ galaxies except ESO055-G013 have
flatter gradients than the average \citet{Sanchez2014} gradient, yet all \hix\
galaxies are within their 1$\sigma$ scatter. However, in the future we expect
larger surveys, such as MaNGA or SAMI paired
with dedicated \hi\ observations to find an answer to this question.

The chemical evolution model of \citet{Ho2015} and \citet{Kudritzki2015} is
able to reproduce measured metallicity gradients. This model suggests that the
local metallicity is mostly dependent on the local stellar to total (\hi\ +
H$_2$) gas mass ratio. This has been observationally confirmed by
\citet{Barrera-Ballesteros2018}. At the galactocentric radii where gas-phase
metallicities can be measured, both the \hi\ and the stellar mass column
densities in \hix\ galaxies are similar to those of control galaxies. It is
only at larger radii, that the \hi\ column densities in control galaxies (or
generally less \hi-rich galaxies) are smaller than in \hix\ galaxies, simply
because the \hix\ \hi\ discs extend much further out. The stellar surface
densities on the other hand are similar at all radii in \hix\ and the
comparison galaxies (similar stellar effective radii; \citealp{Lutz2018}). This
means that the pace of radial metallicity variation in \hix\ and comparison
galaxies should be the same, and this is indeed what we observe. What is not
observed in \hix\ galaxies are large drops in metallicity or star forming
regions with a very low metallicity compared to their surroundings. Thus, major
accretion events are not likely \citep{Moran2012, SanchezAlmeida2014}

The analysis presented in previous papers \citep{Lutz2018,Lutz2017} suggests
that \hix\ galaxies are \hi-rich because they are able to stabilise their \hi\
disc against star formation due to a higher baryonic specific angular momentum.
Examining the dark matter halo spin and \hi\ content of galaxies simulated with
a semi-analytic model, further showed that galaxies with a large \hi\ disc tend
to reside in dark matter halos with a large spin (i.e. angular momentum).

Based on the general interpretation of the scatter of the
mass--metallicity relation, the new results presented here would suggest that
despite the high baryonic specific angular momentum, metal-poor (accreted) gas
would make its way from the outskirts to the centres of \hix\ galaxies, where
it dilutes the local gas. This would imply that \hix\ galaxies are indeed those
galaxies that recently accreted gas in a stochastic accretion process. However,
this would also imply the following two points. First, their extreme \hi\ mass would
require them to accrete more gas than they contained before because they were selected to host at least 2.5\,times more \hi\ than expected from their
optical properties. Second, the accreted gas must have a very high angular
momentum and must  increase the spin of the dark matter halo as well to comply with
the findings from the semi-analytic models.

While \citet{Stewart2011} and \citet{Stewart2017} find in simulations that gas
accreted in the cold mode would indeed come in with a high angular momentum and
form a large co-rotating cold disc (at least in galaxies at redshift > 1), the
two points made above appear very extreme. Furthermore, \citet{Lara-Lopez2013}
find based on observations of \hi\ content, specific star formation rate and
metallicity that there are hardly any metal-poor, \hi-rich galaxies at
intermediate stellar mass and specific star formation rate. These are, however,
exactly the characteristics of the \hix\ galaxies.

One way to reconcile these findings is provided by the semi-analytic models of
\citet{Boissier2000}. These models investigate the chemical evolution of
galaxies with respect to the circular rotation velocity and the halo spin
parameter. They find that galaxies residing in higher spin halos evolve
more slowly than galaxies in lower spin halos. With a slower evolution comes
less enrichment of the ISM and thus a lower gas-phase metallicity, much like the
effect that makes metallicity gradients universal \citep{Sanchez2014}. These
findings have been verified by SDSS observations \citep{Cervantes-Sodi2009}.
\citet{Catinella2015} propose a similar scenario for the most \hi-rich galaxies
detected at redshift 0.2.

Thus, the overall picture for the \hix\ galaxies is the following: they reside
in halos that have higher spins than the halos of average
galaxies in terms of HI content, but not as high  as  necessary to host
low surface brightness galaxies \citep{Boissier2016,Boissier2003,Kim2013}. The
high spin allows them to host larger \hi\ discs than average. At the same time
the high spin slows their evolution such that their ISM is (not yet) as
enriched as it is in galaxies, which reside in average spin halos.

\begin{acknowledgements}
We would like to thank the anonymous referee for helpful comments that improved
the paper.

KL would like to thank Nicola Pastorello for helpful advice on the \wifes\ data
reduction, Natasha Maddox for fruitful discussions, the technical staff at
Siding Springs Observatory for their support during the observations, and the
ANU service desk team for help with the \wifes\ data archive.

Parts of this research were supported by the Australian Research Council Centre
of Excellence for All Sky Astrophysics in 3 Dimensions (ASTRO 3D), through
project number CE170100013.

Besides services and tools already mentioned, data used in this paper were
provided by and analysed with TOPCAT
\citep{Taylor2005}, VizieR\footnote{\url{http://vizier.u-strasbg.fr/}}
\citep{Ochsenbein2000}, AAO Data Central\footnote{\url{datacentral.org.au}},
astropy \citep{AstropyCollaboration2013}, astroquery \citep{Ginsburg2019}, and
matplotlib \citep{Hunter2007}.

\end{acknowledgements}



\bibliographystyle{aa}
\bibliography{library_final.bib}



\begin{appendix} 
\section{Results for individual {\small HIX} galaxies}
\label{app:op_spec}
\begin{figure*}
    \centering
    \includegraphics[width=6.3in]{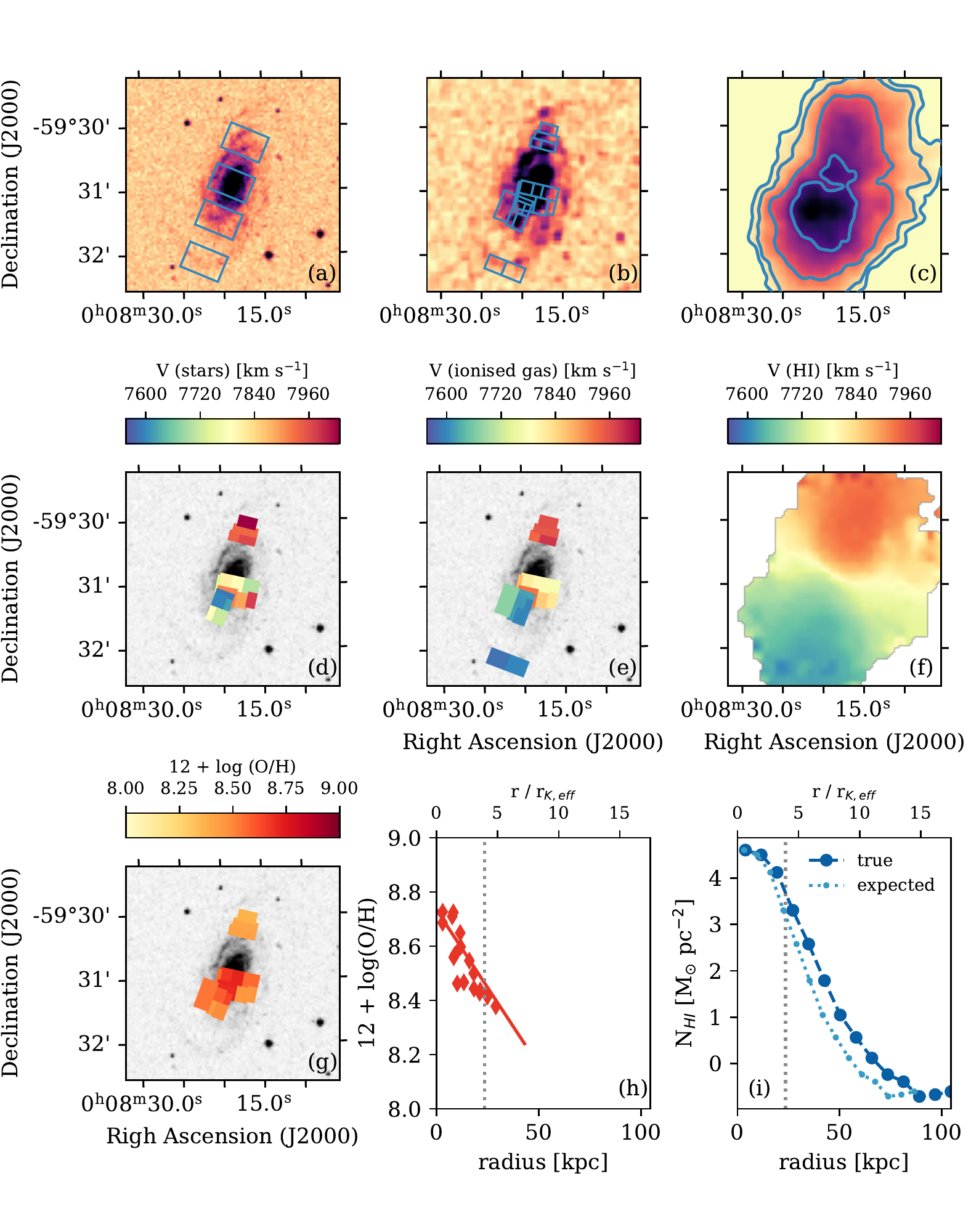}
    \caption{ESO111-G014: \textbf{(a):} Pointings are
    overlaid on
    SuperCOSMOS $B_j$-band images. \textbf{(b):} Star forming regions
    marked on a GALEX $FUV$ images (again $B_j$-band image
    for ESO208-G026). \textbf{(c):} \hi\ moment 0 (intensity) map.
    Contours are at 0.5, 1.0, 2.5, 5.0, 7.5 \msunpcsq. \textbf{(d):}
    Stellar kinematics of single star forming regions overlaid on $B_j$-band
    image. \textbf{(e):} \halpha\ recession velocities of single star
    forming regions overlaid on $B_j$-band image. \textbf{(f):} \hi\
    moment 1 (velocity) map. \textbf{(g):} Metallicity measurements in
    single star forming regions overlaid on $B_j$-band image. \textbf{(h):}
    Radial profile of metallicities with linear fit to the profile, whose
    slope  is the metallicity gradient. The grey dashed line marks the
    25\magasec\ isophotal radius. \textbf{(i):} Radial profile of \hi\
    column density. Dark blue circles and dashed line show the measured values
    corrected for inclination. Light blue points and dotted line show the
    profile rescaled to a 1\msunpcsq\ isophotal radius if the galaxy had
    average \hi\ content (for more details see text). The grey dashed line
     again indicates the 25\magasec\ isophotal radius. \textbf{(a) to (g):}
    All seven panels have the same spatial size; north is up and east is left.
    \textbf{(d) to (e):} Colour scale of the velocities is the same
    in all three panels.}
    \label{fig:eso111_appendix}
\end{figure*}

\begin{figure*}
    \centering
    \includegraphics[width=6.3in]{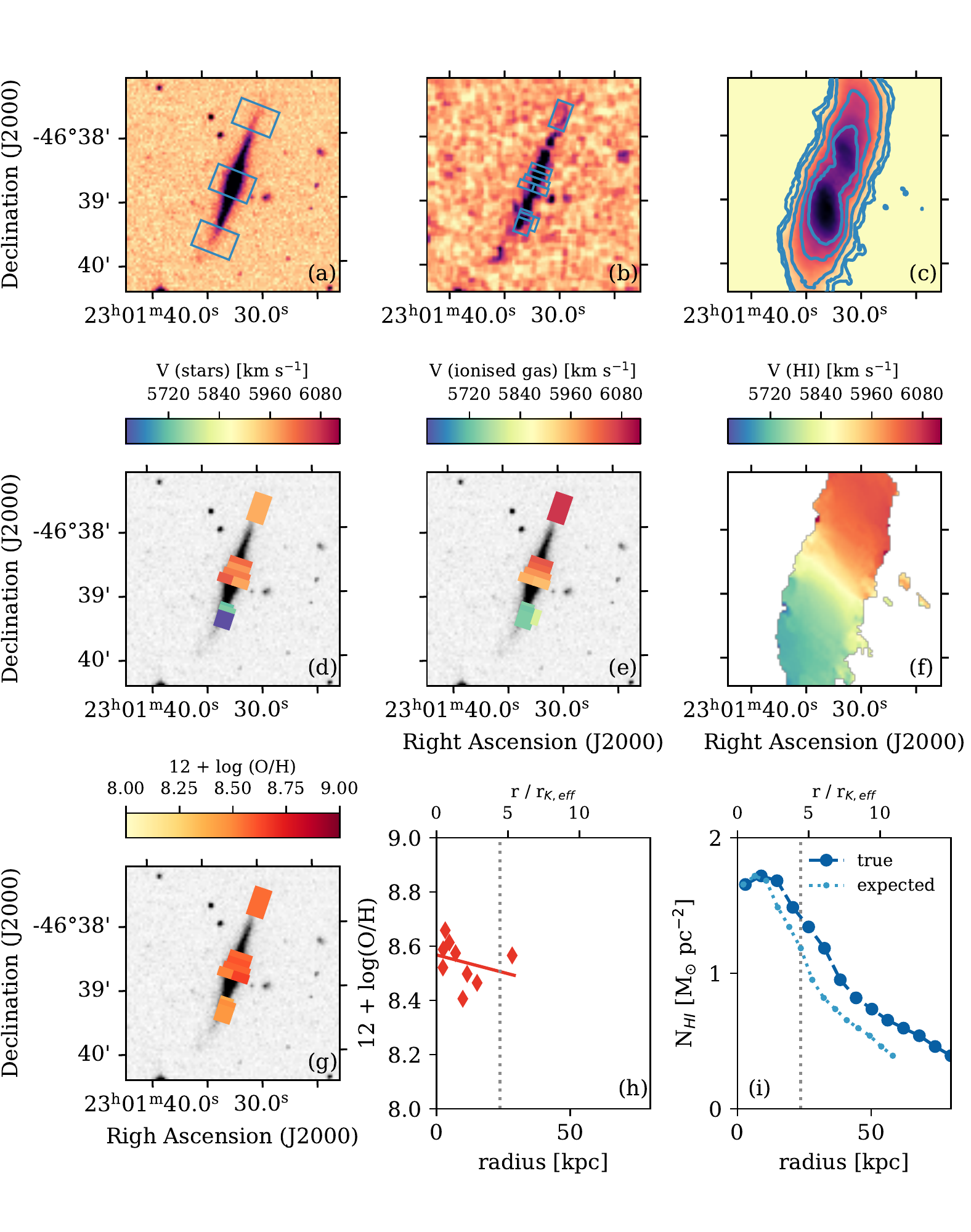}
    \caption{ESO290-G035. Panels as in Fig.~\ref{fig:eso111_appendix}.}
    \label{fig:eso290_appendix}
\end{figure*}

Figures~\ref{fig:eso111_appendix} to \ref{fig:eso290_appendix} show
detailed results for each of the \hix\ galaxies individually. The
maps of stellar and ionised gas kinematics, as well as the maps of metallicity
(Panels (d), (e), and (g)), were obtained with the following procedure: for each
star forming region (their borders are shown in Panel (b)) we derived one
measurement of recession velocity or metallicity. We thus assign this
measurement to the entire star forming region.

The rescaled profile in panel (i) was obtained in the following way.
First the \hi\ mass expected from the \citet{Denes2014} relation was
calculated. Then the expected \hi\ radius (R$_{HI, exp}$, a 1\msunpcsq\
isophotal radius) was calculated from the relation by \citet{Wang2016}. Then
the radii of the profile were rescaled with a factor R$_{HI, exp}$ / R$_{HI,
mea}$ , where R$_{HI, mea}$ is the measured \hi\ radius from \citet{Lutz2018}.

\begin{figure*}
    \centering
    \includegraphics[width=6.3in]{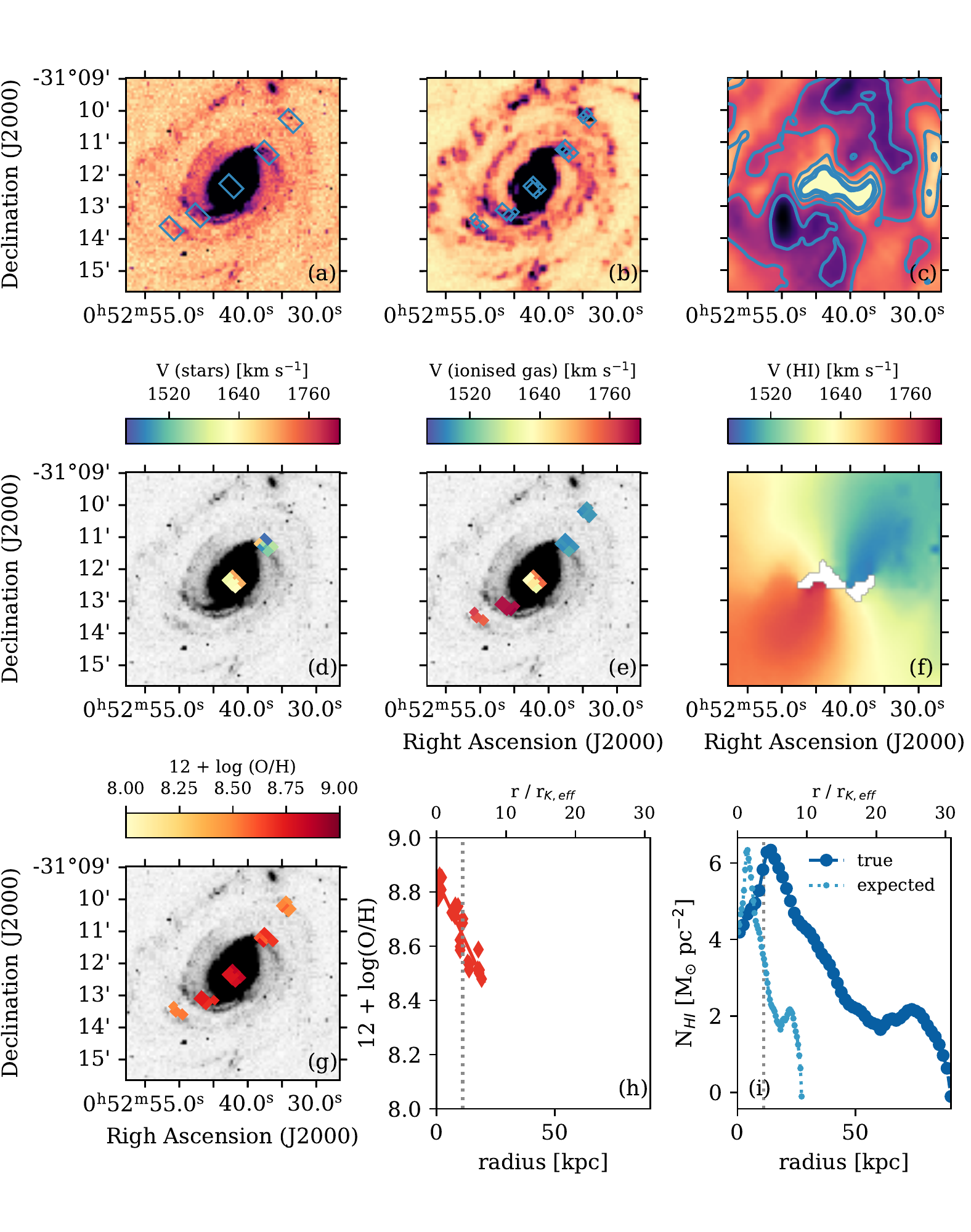}
    \caption{NGC289. Panels as in Fig.~\ref{fig:eso111_appendix}.}
\end{figure*}

\begin{figure*}
    \centering
    \includegraphics[width=6.3in]{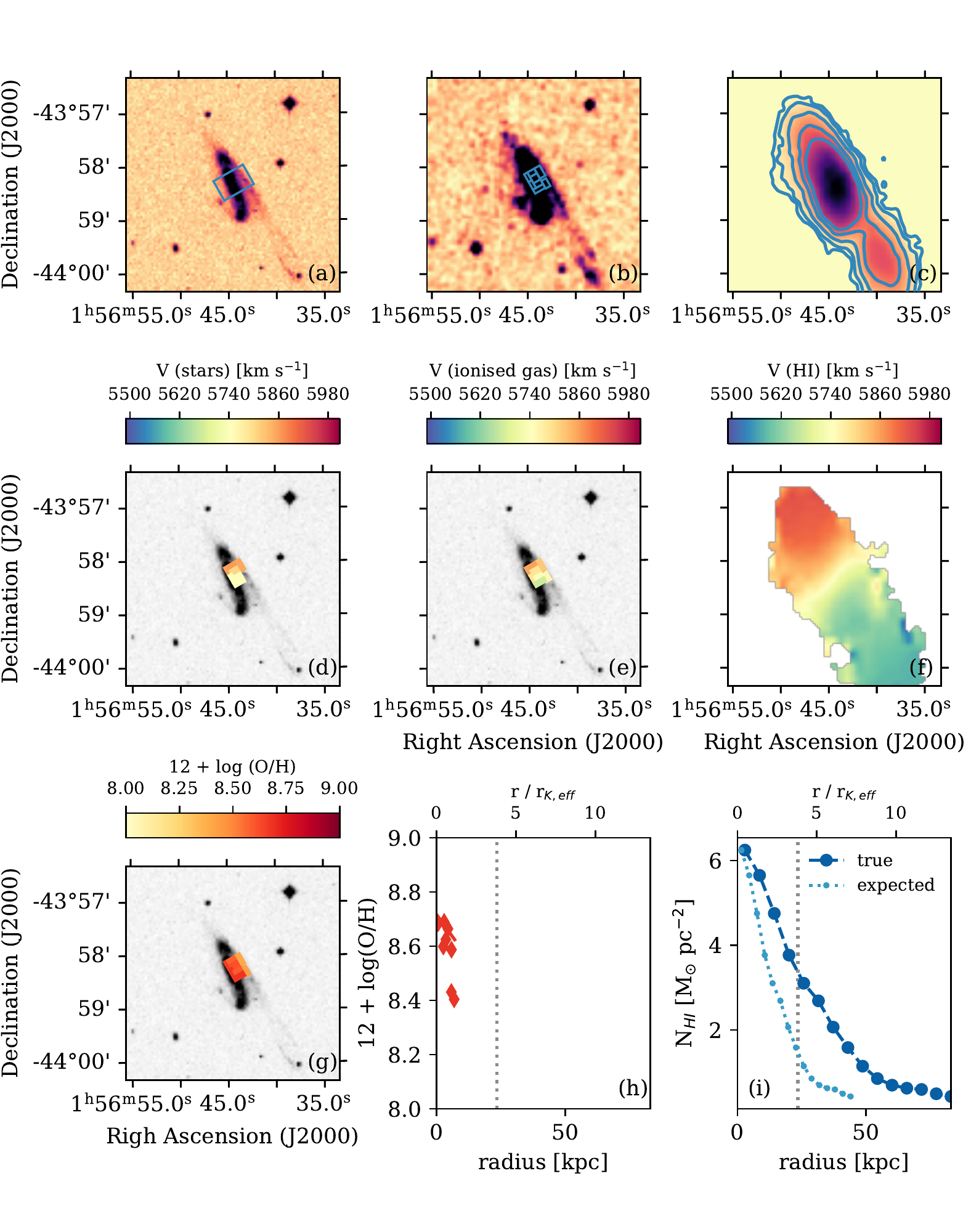}
    \caption{ESO245-G010. Panels as in Fig.~\ref{fig:eso111_appendix}.}
\end{figure*}

\begin{figure*}
    \centering
    \includegraphics[width=6.3in]{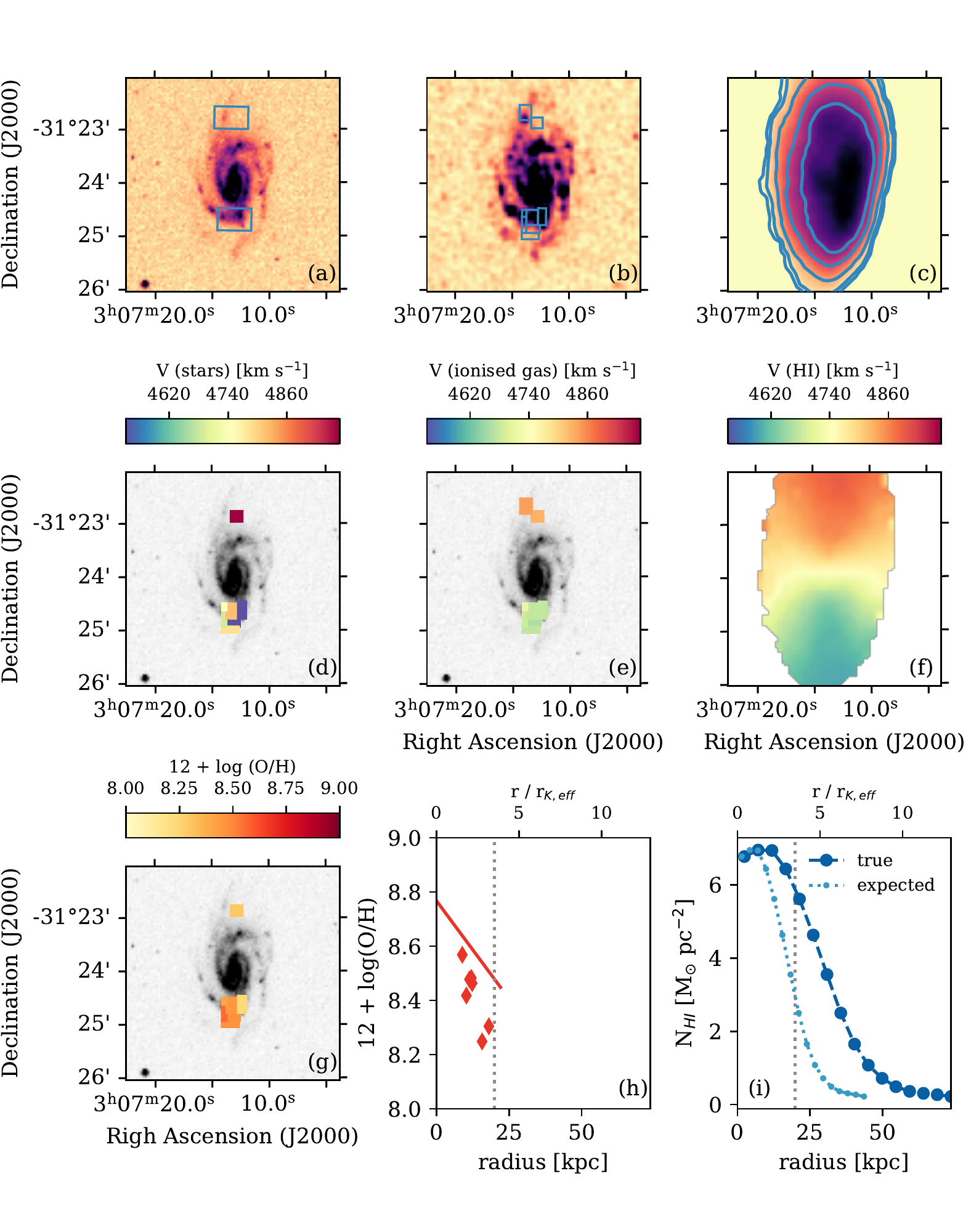}
    \caption{ESO417-G018. Panels as in Fig.~\ref{fig:eso111_appendix}.}
\end{figure*}

\begin{figure*}
    \centering
    \includegraphics[width=6.3in]{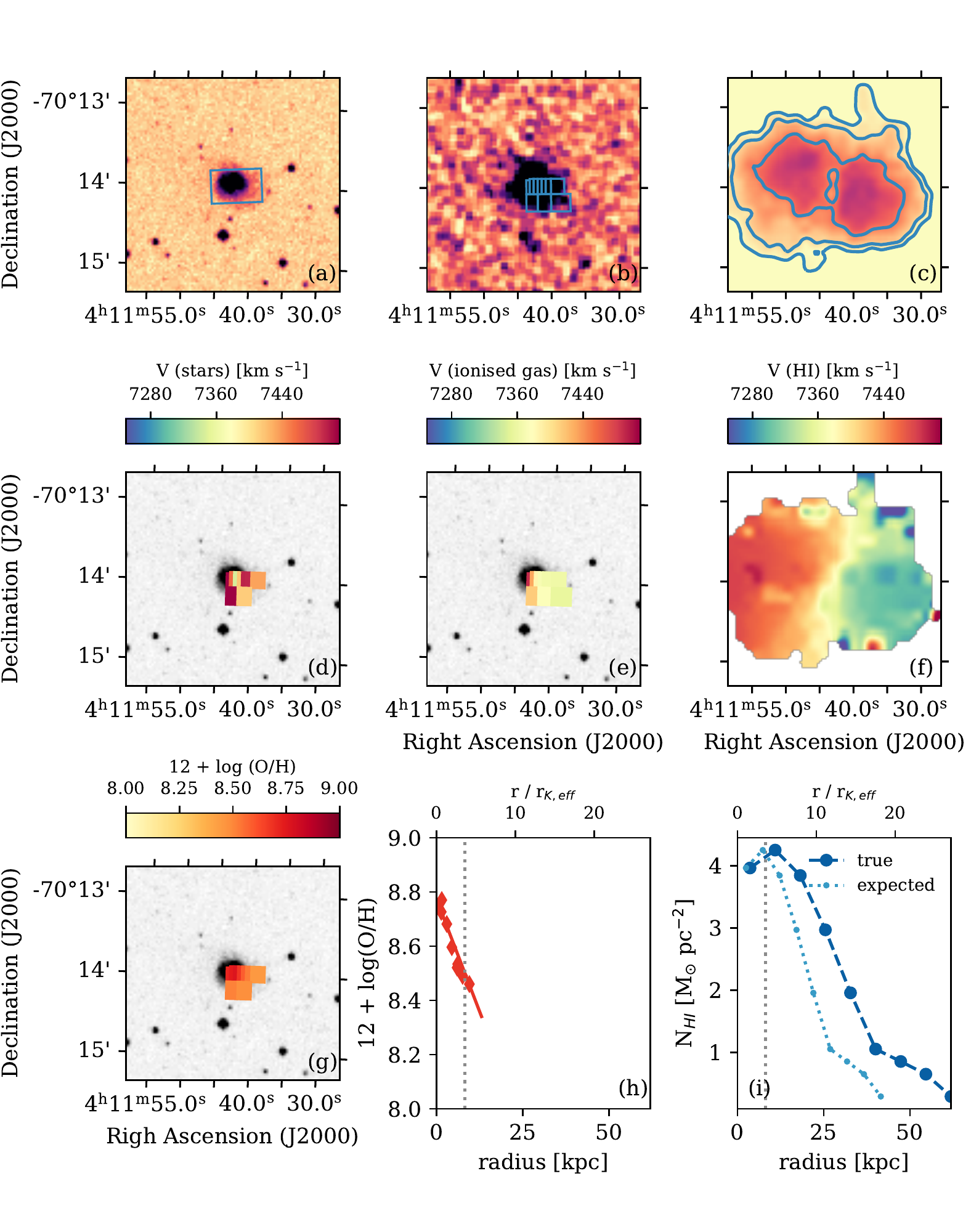}
    \caption{ESO055-G013. Panels as in Fig.~\ref{fig:eso111_appendix}.}
\end{figure*}

\begin{figure*}
    \centering
    \includegraphics[width=6.3in]{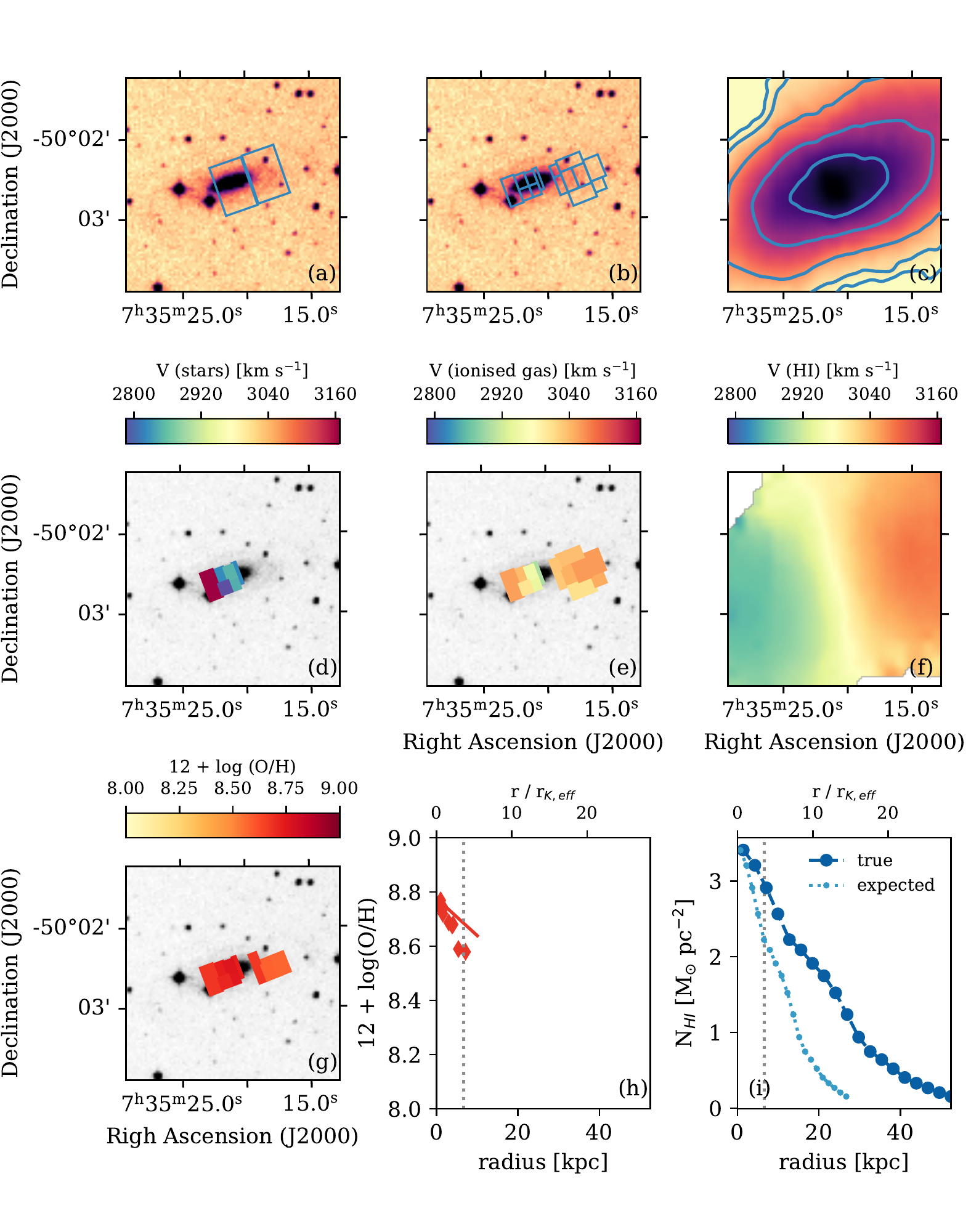}
    \caption{ESO208-G026. Panels as in Fig.~\ref{fig:eso111_appendix}.}
\end{figure*}

\begin{figure*}
    \centering
    \includegraphics[width=6.3in]{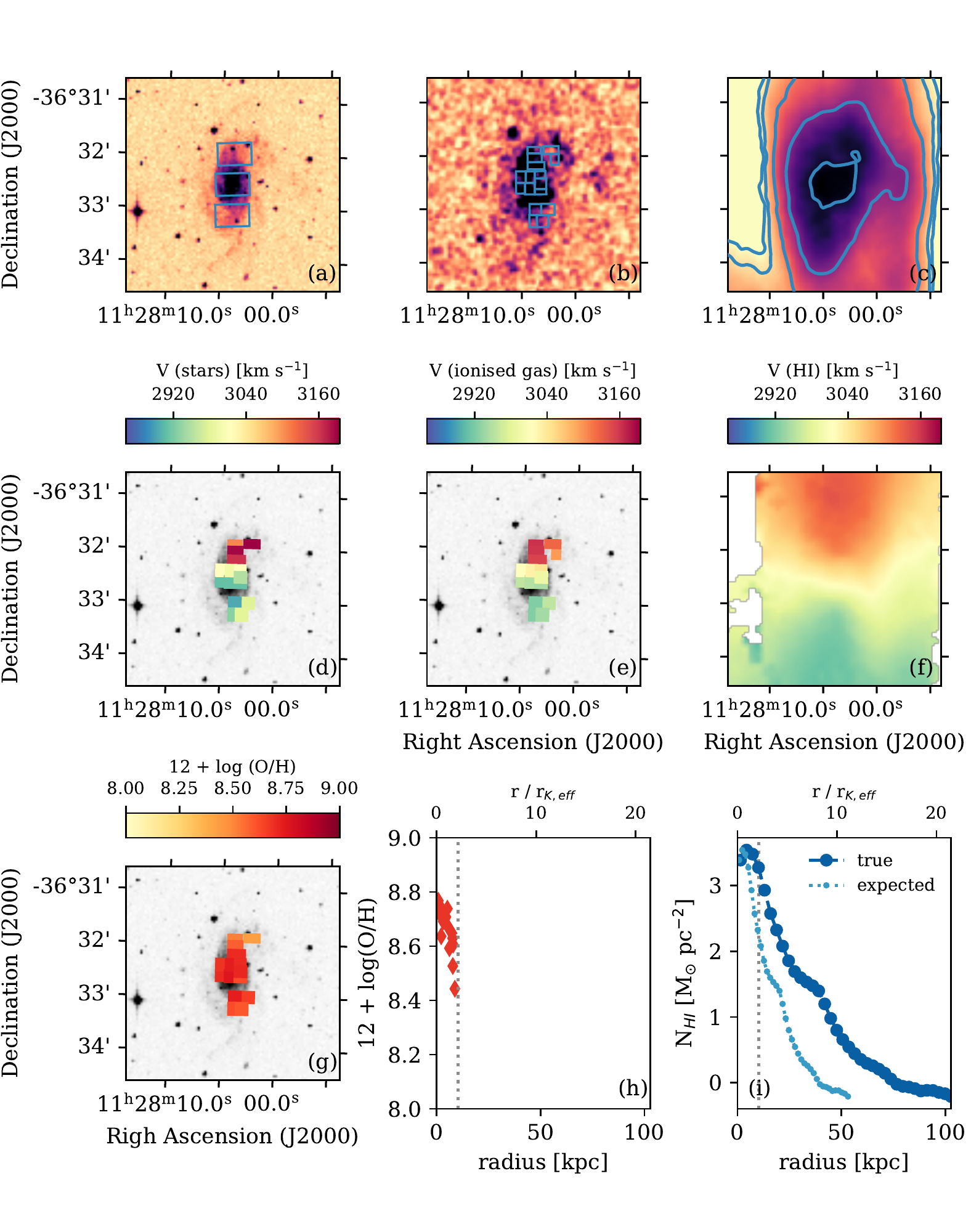}
    \caption{ESO378-G003. Panels as in Fig.~\ref{fig:eso111_appendix}.}
\end{figure*}

\begin{figure*}
    \centering
    \includegraphics[width=6.3in]{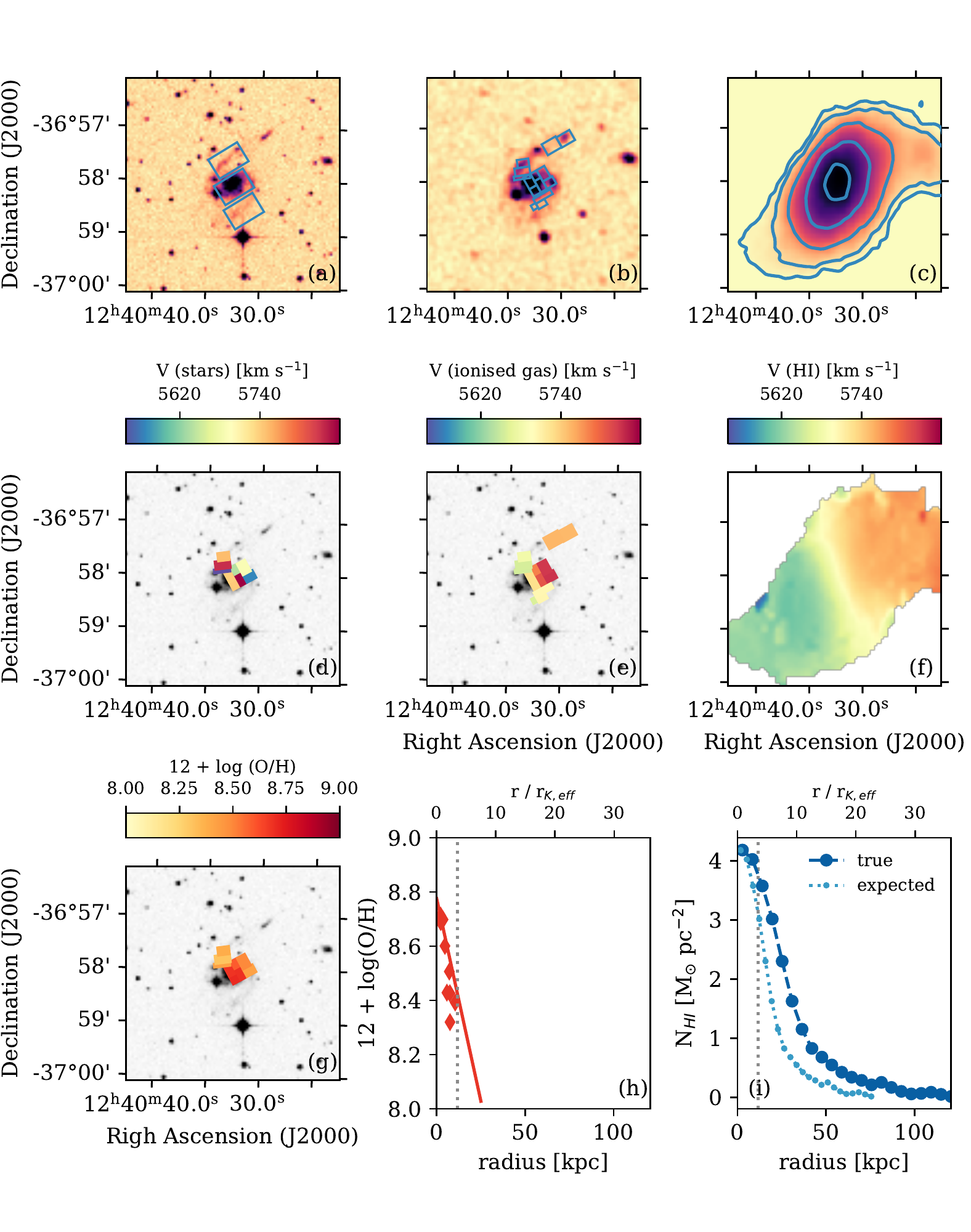}
    \caption{ESO381-G005. Panels as in Fig.~\ref{fig:eso111_appendix}.}
\end{figure*}

\begin{figure*}
    \centering
    \includegraphics[width=6.3in]{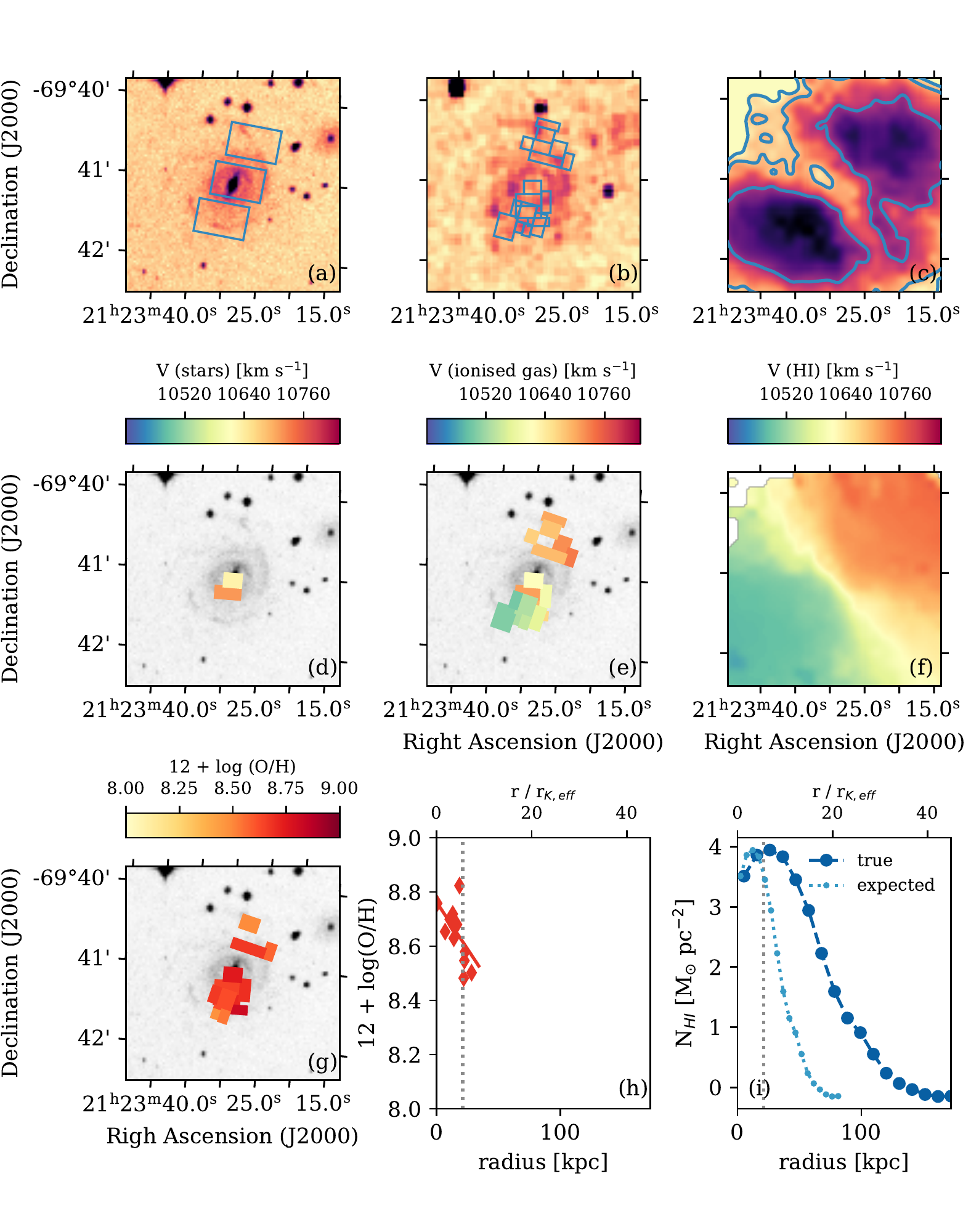}
    \caption{ESO075-G006. Panels as in Fig.~\ref{fig:eso111_appendix}.}
\end{figure*}

\section{Results for IC\,4857}
\begin{figure*}
    \centering
    \includegraphics[width=6.3in]{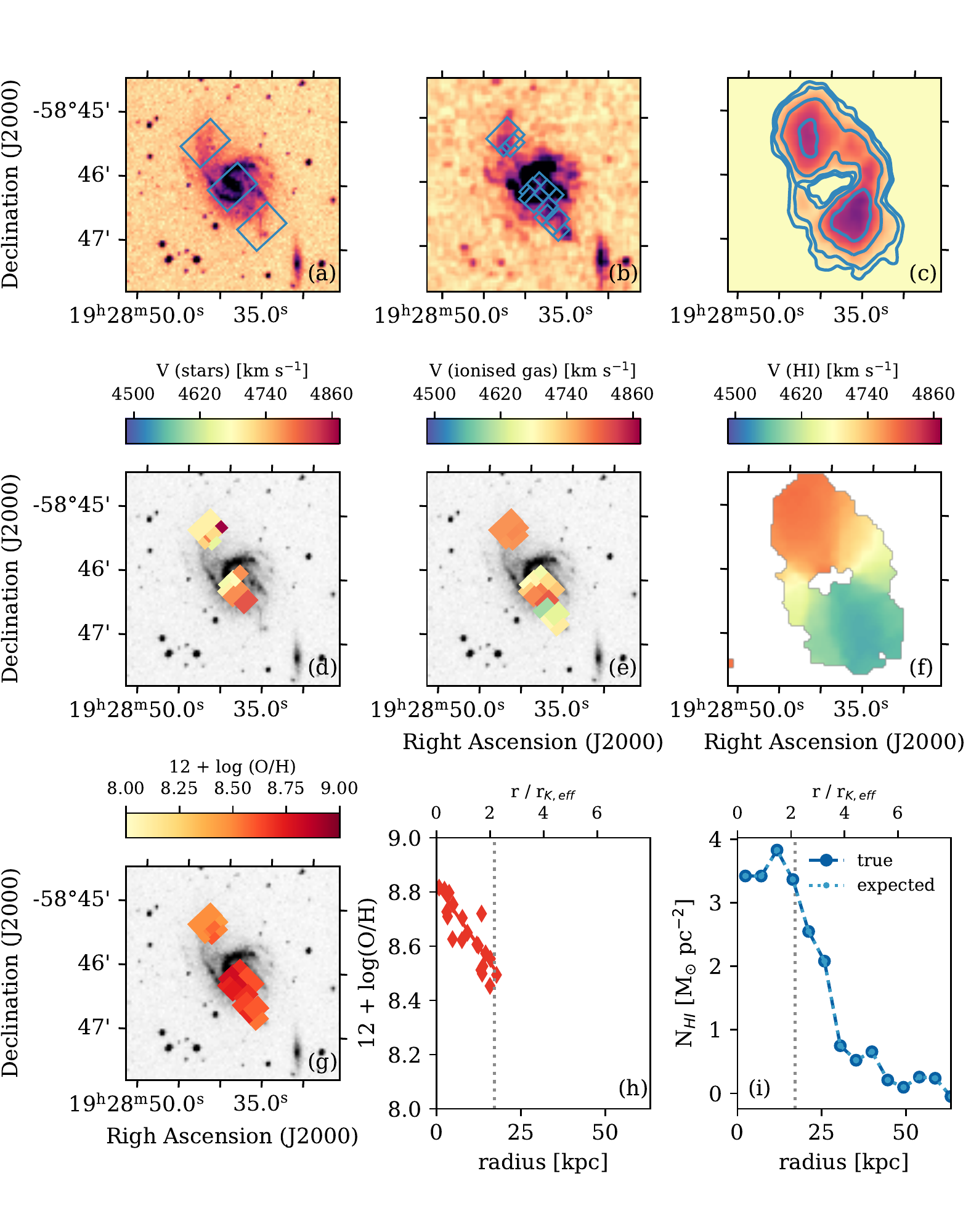}
    \caption{IC\,4857. Panels as in Fig.~\ref{fig:eso111_appendix}.}
    \label{fig:ic4857_appendix}
\end{figure*}

Figure~\ref{fig:ic4857_appendix} shows the detailed results of IC\,4857,
which is the only \hix-control galaxy that was observed with \wifes.
Figure~\ref{fig:ic4857_appendix} shows the same data as
Figs.~\ref{fig:eso111_appendix} to \ref{fig:eso290_appendix}.

\end{appendix}

\end{document}